\documentclass[a4paper,11pt]{article}
\usepackage{jheppub}
\usepackage[T1]{fontenc}
\usepackage{amsthm}
\usepackage{graphicx}
\newtheorem{theorem}{Theorem}
\newtheorem{lemma}[theorem]{Lemma}
\newtheorem{corollary}[theorem]{Corollary}
\newtheorem{proposition}[theorem]{Proposition}
\newcommand{\Tr}{\operatorname{Tr}}
\newcommand{\tr}{\operatorname{tr}}
\newcommand{\Res}{\operatorname{res}}
\newcommand{\FP}{\operatorname{FP}}

\numberwithin{equation}{section}

\begin{document}

\title{Matching lattice and continuum chiral determinants at finite volume}
\author{Zhengwen Liu}
\affiliation{School of Physics \& Shing-Tung Yau Center, Southeast University, Nanjing 210096, China}
\emailAdd{zhengwen.liu@seu.edu.cn}

\abstract{Recent formulations of lattice chiral gauge theories define a complex fermion weight either through Craig's gauge-invariant overlap measure or, following Hsu and Witten, through a continuum Weyl determinant on a reconstructed gauge connection, with its phase fixed by the five-dimensional $\eta$-invariant. For one $SU(5)$ generation on a flat four-torus, we prove that their relative weights agree in the continuum limit along smooth gapped paths, after fixing a normalization and subtracting a local Yang-Mills counterterm, whose coefficient we determine. On bounded smooth families a uniform Dirac gap suffices: the vacuum expansion need not converge. We obtain explicit uniform convergence rates for the logarithmic variations and, for Craig's interpolation, prove the reconstruction bounds that the matching requires, extending Hsu's small-field comparison to finite amplitude. For Hsu's reflection-positive reconstruction with flat collars, the phase variations still match, while the logarithmic coefficient of the quadratic counterterm is enhanced by a calculable, profile-dependent factor. In a hypercharge background on which the vacuum expansion diverges, numerical comparisons at the input gauge field support quadratic convergence of the phase, whereas the modulus acquires a logarithmic correction to quadratic scaling, whose coefficient follows analytically from the free overlap vacuum polarization.}
\maketitle

\section{Introduction}
\label{sec:intro}

Two recent constructions assign a complex fermion weight to a lattice gauge field. In the first, Craig defines a gauge-invariant measure for overlap Weyl fermions \cite{Craig:2026soq}; in the second, Hsu and Witten couple continuum Weyl fermions to a connection reconstructed from the links, fixing the phase by a five-dimensional $\eta$-invariant \cite{Hsu:2026gmi,Witten:2026ezr}. Do the two prescriptions assign the same relative weights in the continuum limit at fixed physical volume, once a local counterterm and a reference normalization have been fixed?

Anomaly matching and closed-loop phases alone do not answer it. Multiplication by $e^{if[A]}$, for a single-valued gauge-invariant functional $f$, leaves the phase around a closed loop unchanged but shifts the relative phase between two backgrounds. We therefore compare how the two weights respond to a variation $b$ of the gauge potential:
\begin{align}
 \delta_b\log W=\delta_b\log|W|+i\,\delta_b\arg W.
\end{align}
Starting with a smooth potential $A$, we form exact link transporters $U_a(A)$, compare the overlap response with the continuum response at $A$, and then replace $A$ by the reconstructed potential $A^r$. This last step is subtle: reconstruction generates Fourier modes at the lattice scale and its frame depends on the background, so both $A^r$ and its variation must be controlled.

Hsu proves convergence on compact finite-dimensional families in a small ball in the weighted Fourier norm $\mathcal W_s$, $s>5$, where the vacuum expansion converges (proposition\,4a in app.\,D.2 in \cite{Hsu:2026gmi}).
His Proposition~4b includes reconstruction under mixed derivative bounds on the reconstructed potential and its field derivative (equation~(D.12) in \cite{Hsu:2026gmi}). The local counterterms, however, remain uncomputed, and the bound $\eta(a)+Ca^2(1+|\log a|)+Ca^6$, with $\eta(a)\to0$, does not fix the overall convergence rate of the comparison (equation~(D.13) in \cite{Hsu:2026gmi}). We use his resolvent formulation, finite-volume Taylor subtractions and continuum determinant factorization, together with Craig's measure estimates: our contribution is the uniform quantitative comparison beyond this small ball, including the actual reconstruction, specified counterterms and their errors.

We work on a flat four-torus of side $L$, with a trivial $SU(5)$ bundle, fermions in $R=\mathbf{10}\oplus\overline{\mathbf5}$, and no Yukawa couplings. The gauge field is prescribed throughout. Our signs refer to $\log W$; the fermion effective action is $-\log W$. Our main results are fourfold: the first two establish the comparison itself, and the last two concern the reconstructions that enter the continuum weight.
\begin{itemize}
\item A uniform Dirac gap alone suffices: convergence of the vacuum expansion is not required. The phase variations converge uniformly when the Fourier amplitudes of the backgrounds and variations have bounded absolute sums and uniformly vanishing tails (theorem~\ref{thm:main}).
\item For amplitudes bounded with weight $(1+|h|)^3$, the full logarithmic variations agree to $O(a/L)$ after a single Yang-Mills counterterm (theorem~\ref{thm:finite}). Its coefficient, including the finite part, follows from the overlap vacuum polarization \cite{Alexandrou:1999wr} and the continuum spectral normalization, and is isotropic for every spin structure with an antiperiodic direction.
\item For Craig's reconstruction, we prove the needed bounds in a periodic frame. The powers of the lattice spacing in these bounds are optimal on families with uniformly bounded input derivatives through second order (appendix~\ref{app:rec}). Gauge invariance then gives the conclusion of Hsu's Proposition~4b without its extra reconstruction hypothesis.
\item With Hsu's flat-collar reconstruction the phase variations still agree, but the logarithmic coefficient of quadratic counterterm picks up a calculable profile factor $\kappa$ (theorem~\ref{thm:collar}). This addresses the quadratic matching problem posed in \cite{Hsu:2026gmi} (app.\,F.5).
\end{itemize}

The proof separates the convergent part of the fermion loop from its low-order amplitudes. A finite resolvent identity absorbs all higher powers of the background into a single bounded inverse. The two-, three- and four-point differences are local up to controlled remainders; the Ward identities then fix their subtraction in terms of the two-point coefficient. For the phase, the first possible contribution has five external insertions, counting the variation as one. Inversion of all four coordinates cancels its zero-momentum term and removes a logarithmic loss. The same estimates yield a growing-momentum window that is sharp when the continuum and plaquette counterterms are used simultaneously.

For every fixed order at fixed external momenta, the error improves to $O(\varepsilon^2\ell_\varepsilon)$, where $\varepsilon=a/L$ and $\ell_\varepsilon=1+\log(1/\varepsilon)$ (appendix~\ref{app:fixedfive}). This fixed-order estimate, however, does not by itself establish the same rate for the full response summed over all orders at a fixed background. Numerical responses evaluated at the original field support faster phase convergence, and the analytic two-point calculation gives a nonzero $\varepsilon^2\log\varepsilon$ correction to the modulus, including its dependence on the momentum direction (section~\ref{sec:numerics}).

Several lines of work come close to parts of this comparison. The continuum phase is described by determinant-line transport and the $\eta$-invariant \cite{DellaPietra:1986qd,Dai:1994kq,Witten:2019bou}. A formal Abelian lattice derivation appears in \cite{Suzuki:1999qw}, and domain-wall constructions relate the phase to a five-dimensional lattice determinant and a local interpolation term \cite{Tatsumi:1999jlf,Kikukawa:2001mw}. Perturbative overlap amplitudes reproduce continuum fermion loops at long wavelengths \cite{Randjbar-Daemi:1995zyu,Randjbar-Daemi:1995lhx}; interpolated gauge fields and local counterterms also underlie the two-cutoff constructions of \cite{Hernandez:1995jc,Hernandez:1996rd,Bodwin:1995rd}. Craig's measure builds on the Ginsparg-Wilson relation, the overlap operator and L\"uscher's solution of the measure problem in perturbation theory \cite{Ginsparg:1981bj,Narayanan:1993sk,Neuberger:1997fp,Luscher:1998du,Luscher:1999un,Luscher:2000zd}. The estimates also use overlap locality \cite{Hernandez:1998et} and one-loop lattice power counting \cite{Adams:2007gh,Capitani:2002mp}.

The paper is organized as follows. Sections~\ref{sec:result}-\ref{sec:match} establish the phase comparison. Sections~\ref{sec:two} and \ref{sec:ward} determine the counterterm and match the complete response; section~\ref{sec:collar} treats the collar reconstruction. Section~\ref{sec:numerics} gives the numerical and analytic tests of the rates. The longer proofs are collected in the appendices.

\section{Fermion weights and phase matching}
\label{sec:result}

The comparison requires compatible chirality, orientation and reference phase. Fix a smooth periodic frame for the trivial bundle on the four-torus. The spin structure is antiperiodic in at least one direction, excluding a free fermion zero mode. In coordinates measured in units of $L$, we write $\varepsilon=a/L=1/N$ and let $A$ be the anti-Hermitian gauge potential, $b=\delta A$ its variation, and $R_*$ the gauge Lie algebra representation. The Dirac operator and free propagator are
\begin{equation}
 D_A=D_0+V(A),\qquad D_0=\gamma^\mu\partial_\mu,\qquad
 V(A)=\gamma^\mu R_*(A_\mu),\qquad S=D_0^{-1}.
 \label{eq:DA}
\end{equation}
The Euclidean gamma matrices are Hermitian. Our chirality matrix and projectors are $\gamma_5=\gamma_1\gamma_2\gamma_3\gamma_4$ and $P_\pm=(1\pm\gamma_5)/2$. A uniform Dirac gap means a positive lower bound on the smallest singular value of $D_A$, or equivalently an upper bound on $\|D_A^{-1}\|$.

The same links $U=U_a(A)$ enter two fermion path integrals. Our convention assigns to each positive link the exact parallel transporter from $x+\varepsilon\hat\mu$ to $x$. The lattice weight $W_{\mathrm{ov}}^{(a)}[U]$ is the fermion path integral defined by the massless overlap Dirac operator (Wilson mass parameter one) and Craig's chiral fermion measure \cite{Craig:2026soq}. For the continuum weight $W_{\mathrm{CW}}^{(a)}[U]$, we first reconstruct a connection $A^r=A_{\rm rec}[U]$ from the links. We use Craig's interpolation functions (app.\,A in \cite{Craig:2026soq}): the resulting reconstruction reproduces affine gauge potentials at linear order, as shown in appendix~\ref{app:rec}. For the reflection-positivity argument (app.\,F.5.1 in \cite{Hsu:2026gmi}), Hsu uses interpolation functions that are constant near each vertex; section~\ref{sec:collar} treats that reconstruction.

The continuum determinant is a section of a complex line bundle over gauge-field space. Its Quillen metric gives the modulus \cite{Quillen:1985}. Hsu fixes a unit vector at a reference connection and transports it with the Bismut-Freed (BF) connection; the weight is the coefficient of the determinant section in this parallel frame \cite{Hsu:2026gmi}. Restricting $R$ to $S(U(3)\times U(2))$ gives one Standard Model generation; additional identical generations multiply the weights and add their logarithmic variations.

To identify the two Weyl operators, fix Craig's ordered coordinates $(x^1,x^2,x^3,x^4)$, put $t=x^4$, and use the Pauli basis $\gamma_j=\sigma_1\otimes\sigma_j$, $\gamma_4=\sigma_2\otimes1$ (app.\,B.1 in \cite{Craig:2026soq}). Then $\gamma_5=-\sigma_3\otimes1$, and the domain of the Weyl operator, the negative-chirality subspace, consists of the upper two components. Write $\nabla_\mu=\partial_\mu+R_*(A_\mu)$. After multiplication by the constant $-i$, which leaves logarithmic variations unchanged, Craig's Weyl operator is $P_C[A]=\nabla_t-i\sigma_j\nabla_j$. In Hsu's conventions (app.\,F.5.2 in \cite{Hsu:2026gmi}), it is $P_H[A]=\nabla_t+i\sigma_j\nabla_j=-P_C[A]^\dagger$. Hsu matches the chiralities with the complementary overlap projector (relations~(D.9)-(D.10) in \cite{Hsu:2026gmi}). We take instead the reflection $r(\mathbf x,t)=(-\mathbf x,t)$, which preserves every spin structure since each boundary-condition sign equals its inverse. With $(T_r\psi)(\mathbf x,t)=\psi(-\mathbf x,t)$, $(r^*A)_j=-A_j\circ r$ and $(r^*A)_t=A_t\circ r$,
\begin{equation}
 T_r^{-1}P_H[r^*A]T_r=P_C[A] .
 \label{eq:orientation}
\end{equation}
The same identity holds along $A+ub$, with $b$ pulled back as a one-form. Let $W_{\rm H}(C)$ denote Hsu's continuum weight for a smooth connection $C$ written in his coordinates. For links given in Craig's coordinates the continuum prescription is
\begin{equation}
 W_{\mathrm{CW}}^{(a)}[U]=W_{\rm H}\big(r^*A_{\rm rec}[U]\big) .
 \label{eq:cwdef}
\end{equation}
Equivalently, one may keep the same ordering of the axes and take the complex conjugate of Hsu's relative phase. In either description the phases are normalized at corresponding reference connections.

The five-dimensional description fixes the same phase. The spin structure on an antiperiodic circle extends over a disk, so the torus bounds $Y=T^3\times D^2$, and the trivial bundle extends over $Y$. Choose a product metric and a gauge field of product form near the boundary. When the boundary Dirac operator is invertible, the reduced invariant $\xi_Y=(\eta_{\rm APS}(0)+\dim\ker\mathcal D_Y)/2$ of the five-dimensional operator with Atiyah-Patodi-Singer boundary conditions \cite{Atiyah:1975jf} defines a unit vector in the dual of the boundary determinant line. Witten's phase for positive boundary chirality is $\exp(-2\pi i\xi_Y)$, his $\eta$ being $2\xi_Y$ \cite{Witten:2026ezr}; reversing the orientation, and with it the boundary chirality, conjugates this phase. The covariant variation of the unit vector is the fibre integral of the degree-six anomaly polynomial \cite{Bismut:1986BF2,Dai:1994kq,Yonekura:2016wuc}, which vanishes for $R$, and $\Omega^{\rm Spin}_5(BSU(5))=0$ \cite{Garcia-Etxebarria:2018ajm} excludes a residual global anomaly. The filling therefore gives a parallel phase convention. After normalization at a common reference it agrees with parallel transport by the Bismut-Freed connection (proposition~6 in \cite{Hsu:2026gmi}).

We measure the size and regularity of the gauge field by its Fourier coefficients in the fixed gauge frame. For $s\ge0$ the weighted Wiener norm, which is the Fourier algebra norm (equation~(D.8) in \cite{Hsu:2026gmi}), is
\begin{equation}
 \|A\|_{\mathcal W_s} 
 =\sum_{h\in2\pi\mathbb Z^4}(1+|h|)^s   \sum_{\mu=1}^4\|\widehat A_\mu(h)\|.
 \label{eq:wiener}
\end{equation}
We abbreviate these norms by $A_s=\|A\|_{\mathcal W_s}$ and $b_s=\|b\|_{\mathcal W_s}$.
For a smooth field $f$ (a potential or a variation), set
\begin{equation}
 \omega_\varepsilon(f)=\sum_{h\in2\pi\mathbb Z^4}
 \min\{1,\varepsilon(1+|h|)\}\sum_\mu\|\widehat f_\mu(h)\| .
 \label{eq:fouriermodulus}
\end{equation}
At momenta well below the lattice scale each amplitude is weighted by $\varepsilon(1+|h|)$; modes at or above that scale enter with their full amplitude. This distinction allows a phase estimate even when the Fourier coefficients obey no uniform power-law decay.

\begin{theorem}\label{thm:main}
Fix the torus, its spin structure, the representation and finite bounds $B,G$. There are $C,\omega_0,\varepsilon_0>0$ such that for every smooth $A$ with $\|A\|_{\mathcal W_0}\le B$ and $\|D_A^{-1}\|\le G$, both weights on $U_a(A)$ are nonzero if $\varepsilon<\varepsilon_0$ and $\omega_\varepsilon(A)\le\omega_0$. For every smooth variation $b$ with $\|b\|_{\mathcal W_0}<\infty$,
\begin{equation}
 \left|\operatorname{Im}\,\delta_b\log
       \frac{W_{\mathrm{ov}}^{(a)}[U_a(A)]}{W_{\mathrm{CW}}^{(a)}[U_a(A)]}\right|
 \le C\bigl(\omega_\varepsilon(b)+\omega_\varepsilon(A)\|b\|_{\mathcal W_0}\bigr).
 \label{eq:phasefine}
\end{equation}
The derivative is taken along $A+tb$ on both sides. In particular, if $0<\alpha\le1$ and $\|A\|_{\mathcal W_\alpha}\le B$, then for $\varepsilon<\varepsilon_0$ with $B\varepsilon^\alpha\le\omega_0$,
\begin{equation}
 \left|\operatorname{Im}\,\delta_b \log \frac{W_{\mathrm{ov}}^{(a)}[U_a(A)]}{W_{\mathrm{CW}}^{(a)}[U_a(A)]}\right|
       \le C_{B,G}\varepsilon^\alpha\|b\|_{\mathcal W_\alpha}.
 \label{eq:main}
\end{equation}
\end{theorem}

Indeed $\min(1,x)\le x^\alpha$, so the constant in \eqref{eq:main} can be chosen independently of $\alpha$; it is the spacing threshold that deteriorates as $\alpha\downarrow0$. In four dimensions $\|f\|_{\mathcal W_\alpha}\le C_{s,\alpha}\|f\|_{H^s}$ for $s>2+\alpha$, so $s>3$ suffices for the endpoint rate $O(\varepsilon)$. More generally, consider bounded sets of smooth backgrounds and variations in $\mathcal W_0$ whose Fourier tails above $M$ tend to zero uniformly as $M\to\infty$. The contribution from $|h|\le M$ to $\omega_\varepsilon(f)$ is at most $\varepsilon(1+M)\|f\|_{\mathcal W_0}$. Taking first $M$ large, then $\varepsilon$ small, proves uniform convergence of phase variations on such sets if the backgrounds have a common gap; the families need not be bounded in any $\mathcal W_\alpha$ with $\alpha>0$.

\section{Phase variations and ultraviolet convergence}
\label{sec:response}

The phase is governed by ultraviolet-finite loops, which makes the two prescriptions comparable term by term. A free massless propagator falls off as $1/|k|$ at large loop momentum $k$, so in four dimensions a fermion loop with five propagators converges. For the anomaly-free representation $R$ the first possible contribution to the phase variation has five external gauge fields, counting the variation $b$ as one, and the two phase variations can therefore be compared through ordinary operator traces.

\subsection{Logarithmic variation of the overlap weight}

On the lattice the variation of the weight is a resolvent trace with a fixed chiral projector, plus Craig's measure term. For the overlap Dirac operator $D_a=D_a(A)$, measured in units of $L^{-1}$, define
\begin{equation}
 S_a = D_a(0)^{-1},\qquad V_a=D_a(A)-D_a(0),\qquad
 K_a = V_a S_a,\qquad J_a=(\delta_b V_a)S_a .
 \label{eq:latK}
\end{equation}
Craig's variation formula~(37) in \cite{Craig:2026soq}, which we use as stated there, simplifies on the invertible chart. With our choice of Wilson kernel parameter, the field-dependent chiral projector is $\widehat P_-=[1-\gamma_5(1-\varepsilon D_a)]/2$, and the Ginsparg-Wilson intertwining $D_a\widehat P_-=P_+D_a$ replaces the moving projector by the fixed one (relations~(D.14)-(D.15) in \cite{Hsu:2026gmi}). Substituting $D_a^{-1}=S_a(1+K_a)^{-1}$ then yields
\begin{equation}
 \delta_b\log W_{\mathrm{ov}}^{(a)}
 =\Tr[J_a(1+K_a)^{-1}P_+]-i\mathfrak L_b .
 \label{eq:ov}
\end{equation}
Here $\mathfrak L_b$ is the real measure term, Craig's measure current contracted with the infinitesimal link variation (equation~(34) in \cite{Craig:2026soq}). The inverse estimates in section~\ref{sec:conv} and appendix~\ref{app:odd} justify this formula on the backgrounds of theorem~\ref{thm:main}.

To isolate the terms with at most four propagators, we use the finite resolvent identity
\begin{equation}
 (1+K)^{-1}=\sum_{j=0}^3(-K)^j+K^4(1+K)^{-1}.
 \label{eq:split}
\end{equation}
The dynamical phase vanishes through fourth order in the external fields. Let $\mathcal C$ denote complex conjugation followed by the spin matrix specified by equation~(175) in \cite{Craig:2026soq}. This antiunitary map preserves the gamma matrices and $P_+$, and intertwines the Wilson and overlap operators in $R$ and $R^*$. On real fields, corresponding traces are complex conjugates. Their degree-$n$ coefficients contain ordered colour traces of $n$ generators, which agree in $R$ and $R^*$ for $n\le4$ (relations~(78)-(81) in \cite{Craig:2026soq}). For the representation used here, this follows from $\chi_R(g)-\chi_{R^*}(g)=\det(1-g)$, whose expansion starts at fifth order, together with the Lie algebra commutators that reduce ordered traces. Hence these coefficients are real. Real and imaginary parts are taken on real fields before multilinear extension to complex Fourier modes, as in \eqref{eq:reallegs}.

A single lattice vertex can have several external gauge-field insertions because the link variable depends nonlinearly on the gauge potential. We call such vertices contact vertices; their ordered expansion is given by formulas~(D.16)-(D.20) in \cite{Hsu:2026gmi}. Even the polynomial in \eqref{eq:split} can therefore contribute at fifth and higher orders in the external fields. Appendix~\ref{app:vertices} bounds the sum of these terms by $C_B\varepsilon(1+|\log\varepsilon|)\|b\|_\infty$ and the measure term by $C_B\varepsilon\|b\|_\infty$. At $\alpha=1$, appendix~\ref{app:odd} combines the five-point contact terms with the five-propagator loop before applying inversion in all four coordinates. This removes the logarithm from the comparison. Up to the errors just estimated, the lattice phase variation is $\operatorname{Im}\mathcal T(K_a,J_a)$, where
\begin{equation}
 \mathcal T(K,J)=\Tr[JK^4(1+K)^{-1}P_+] .
 \label{eq:tail}
\end{equation}

\subsection{Logarithmic variation of the continuum weight}

The continuum variation takes the same form once its zeta-regularized trace is split into a local residue and a trace-class remainder. In Craig's coordinates the continuum weight is $W_{\rm cont}(A)=W_{\rm H}(r^*A)$ by \eqref{eq:cwdef}. Let $P=P_0+V$ be the Weyl operator from negative- to positive-chirality spinors, with $V$ the corresponding block of $V(A)$, and set $\mu L=1$, where $\mu$ is the mass scale in zeta regularization. For invertible $P$ and $P_0$, $Q=P^\dagger P$ and $Q_0=P_0^\dagger P_0$ are positive elliptic operators, and $\tr^Q(T)=\FP_{z=0}\Tr(TQ^{-z})$ is the constant term in the Laurent expansion at $z=0$. With the Bismut-Freed spectral cutoff below the smallest eigenvalue of $Q$, the logarithmic variation in a parallel frame is given by the finite-part formula~(D.36) in \cite{Hsu:2026gmi}, the source-free case of formula~(F.46) in \cite{Hsu:2026gmi}; see also \cite{Bismut:1986BF1,Bismut:1986BF2}:
\begin{equation}
 \delta_b\log W_{\rm cont}(A)=\tr^{Q}(P^{-1}\delta_bP).
 \label{eq:BF}
\end{equation}
Since $Q$ and $Q_0$ have the same principal symbol, the change of weight is local \cite{Wodzicki:1984,Kontsevich:1994xe,Cardona:2002pn}: for $T=P^{-1}\delta_bP$,
\begin{equation}
  \tr^{Q}(T)-\tr^{Q_0}(T)
  =-\tfrac12\Res\!\left[T(\log Q-\log Q_0)\right].
  \label{eq:weight}
\end{equation}
The residue integrates the trace of the homogeneous symbol of degree $-4$ over position and the unit momentum sphere. The symbol of $T$ starts at degree $-1$ and already contains $b$. Each further insertion of the potential, in $P^{-1}$ and in the resolvent expansion of the logarithms, lowers the degree by at least one, since $Q-Q_0=P_0^\dagger V+V^\dagger P_0+V^\dagger V$ has differential order at most one in its linear terms and zero in its quadratic terms, and a derivative acting on a coefficient comes with a momentum derivative. Hence at most four colour generators occur in the residue. The map $\mathcal C$ intertwines the continuum operators and their spectral weights in $R$ and $R^*$, first for large real $z$ and then by analytic continuation of the finite parts; since ordered colour traces agree through degree four, the residue is real.

Set $K_-=P_0^{-1}V$ and $J_-=P_0^{-1}\delta_b V$ on negative-chirality spinors. Applying \eqref{eq:split} to $T=(1+K_-)^{-1}J_-$ gives the four traces $\tr^{Q_0}(K_-^\ell J_-)$, $0\le\ell\le3$, which are real by the same colour argument since $Q_0$ acts trivially on colour, and a trace-class remainder with five free propagators. Moving this remainder to positive chirality with $S_+=P_0^{-1}$, $K_+=VS_+$ and $K_-S_+=S_+K_+$, cyclicity gives
\begin{align}
 &
 \delta_b\arg W_{\rm cont}(A)
   =\operatorname{Im}\Tr_{-}[K_-^4(1+K_-)^{-1}J_-]
   =\operatorname{Im}\,\mathcal T(K,J),
   \nonumber\\
   &
   K=V(A)S,\qquad J=V(b)S .
 \label{eq:cw}
\end{align}
Up to the lattice contact and measure terms estimated above, the two phase variations have the same trace representation \eqref{eq:tail}, with lattice and continuum operators. Before the imaginary part is taken, the first trace in \eqref{eq:cw} is the logarithmic derivative of the regularized Fredholm determinant $\det_5(1+K_-)$ (equations~(D.32)-(D.37) in \cite{Hsu:2026gmi}); related factorizations are developed in \cite{Dang:2019byu}. The real parts of the four regulated traces can depend nonlocally on $A$; they enter the comparison of moduli in section~\ref{sec:modulus}.

\section{Momentum-space estimates}
\label{sec:conv}

The trace comparison uses Schatten bounds on a vertex multiplied by its propagator. We identify the normalized lattice Fourier basis with continuum modes in a half-open first Brillouin zone and extend $S_a,K_a,J_a$ by zero outside it. Then $(1+K_a)^{-1}$ acts as the identity outside the zone, and the embedding preserves $P_+$. For singular values $s_j(X)$, the Schatten norm is $\|X\|_p=(\sum_j s_j(X)^p)^{1/p}$.

\begin{lemma}\label{lem:S}
Let $p>4$, $0<\alpha\le1$ and $\|A\|_{\mathcal W_\alpha}\le B$. For sufficiently small $\varepsilon$,
\begin{align}
 \|K_a-K\|_p &\le C_{B,p}(\varepsilon^{1-4/p}+\varepsilon^\alpha)
                        \|A\|_{\mathcal W_\alpha},\nonumber\\
 \|J_a-J\|_p &\le C_{B,p}(\varepsilon^{1-4/p}+\varepsilon^\alpha)
                        \|b\|_{\mathcal W_\alpha}.
 \label{eq:S5}
\end{align}
The constants can be chosen uniformly for all $p\ge5$.
\end{lemma}
The proof, in appendix~\ref{app:proofs}, compares each linear vertex and its propagator with their continuum limits at momenta below the lattice scale and bounds the remaining modes by the free propagator alone. It also gives the operator-norm bounds \eqref{eq:tameop} and \eqref{eq:modop} used below.

The spectral gap now transfers to the overlap operator. Since $(1+K)^{-1}=1-V(A)D_A^{-1}$, its norm is at most $1+CBG$. By \eqref{eq:modop}, choosing $\omega_\varepsilon(A)$ sufficiently small makes $\|(K_a-K)(1+K)^{-1}\|<1/2$. Thus $1+K_a$ has a uniformly bounded inverse on the Fourier space defined above, and $D_a$ is invertible on the lattice.

Estimating the difference of the two traces \eqref{eq:tail} factor by factor, with one difference in $\mathcal S_p$ and the four other propagator factors in $\mathcal S_t$, $1/p+4/t=1$, gives $O(\varepsilon^\alpha)$ for $0<\alpha<1$ and $O(\varepsilon\ell_\varepsilon)$ at $\alpha=1$, the logarithm coming from the sum of $|k|^{-4}$ over four-dimensional momentum shells up to the lattice scale. Appendix~\ref{app:odd} removes this logarithm by combining the five-point contact terms with the five-propagator loop before applying inversion in all four coordinates.

\section{Reconstruction and phase matching}
\label{sec:match}

The estimates so far compare the overlap with the continuum determinant at the original field $A$. The second prescription instead uses $A^r=A_{\rm rec}[U_a(A)]$. We must therefore control $A^r-A$ and the induced variation $b^r=\partial_u A^r(A+ub)|_{u=0}$, including the changing frame.

Lemma~\ref{lem:reconstruction} supplies a common periodic frame. The linear reconstruction reproduces constant potentials and uses only paths within distance $C\varepsilon$ of the observation point. A unit plane wave of momentum $h$ varies by at most $C'\min(1,\varepsilon|h|)$ on those paths. Summing the linear errors and using the nonlinear bounds in the proof of the lemma gives, for $A_0\le B$,
\begin{equation}
 \|A^r-A\|_\infty\le C_B\omega_\varepsilon(A),\qquad
 \|b^r-b\|_\infty\le C_B\omega_\varepsilon(b).
 \label{eq:modrec}
\end{equation}
Indeed the nonlinear errors are $O_B(\varepsilon A_0)$ and $O_B(\varepsilon b_0)$, respectively. First choose $\varepsilon_0$ small enough at fixed $B$ for the link expansion, the plaquette admissibility condition and the construction of the common frame. Then choose $\omega_0$ small enough at fixed $B,G$ for the overlap inverse bound in section~\ref{sec:conv} and for $\|V(A^r-A)D_A^{-1}\|<1/2$ whenever $\omega_\varepsilon(A)\le\omega_0$. This gives $\|D_{A^r}^{-1}\|\le2G$ and proves nonvanishing on both sides of the comparison.

For smoother backgrounds, the same construction yields a stronger conclusion: in a single periodic frame the reconstruction error and its field derivative have $C^j$ norms of order $\varepsilon^{2-j}$ for $j=0,1,2$, as specified in \eqref{eq:mixedrecon}. These are the powers required by equation~(D.12) in \cite{Hsu:2026gmi}, which Hsu states in a frame obtained by radial transport; since the weight is gauge invariant, the periodic frame suffices (appendix~\ref{app:rec}). At $j=2$ the conclusion is boundedness, not convergence. The estimate concerns the interpolation functions fixed in section~\ref{sec:result}.

The periodic frame differs from Hsu's radial-transport frame by a background-dependent gauge transformation. Its derivative adds a gauge variation to the transformed tangent, as displayed in \eqref{eq:recframes}. The continuum scalar weight is gauge invariant in the common phase convention, so this term has zero response. The estimates in our frame therefore compare the actual weights.

\begin{proof}[Proof of theorem~\ref{thm:main}]
The map $A\mapsto V(A)S$ is continuous from $L^\infty$ to $\mathcal{S}_5$. Replacing the factors of $\mathcal T$ one by one in \eqref{eq:cw} and applying the inverse bounds at $A$ and $A^r$ gives
\begin{equation}
 \big|\delta_{b^r}\arg W_{\rm cont}(A^r)-\delta_b\arg W_{\rm cont}(A)\big|
 \le C_{B,G}\big(\|b^r-b\|_\infty+\|A^r-A\|_\infty\|b\|_\infty\big).
 \label{eq:pullback}
\end{equation}
Thus \eqref{eq:modrec} bounds the cost of reconstruction by the right-hand side of \eqref{eq:phasefine}. It remains to compare the two prescriptions at $A$. The terms with at most four external insertions have zero imaginary part. For the five-point coefficient, lemma~\ref{lem:projectedfive} bounds the error by $C\min(1,\varepsilon H)$, where $H=1+\sum_i|h_i|$. This weight can be assigned to one external leg at a time in the Fourier sum. In the remainder with six propagators, after one factor has been replaced the other five can each be estimated in $\mathcal S_5$. The change in the remaining factor is bounded by \eqref{eq:modop}. Higher-order contact terms and the measure contribute $O_B(\varepsilon b_0)$; the details are given in appendix~\ref{app:odd}. Together these estimates prove \eqref{eq:phasefine}. Using $\omega_\varepsilon(f)\le\varepsilon^\alpha\|f\|_{\mathcal W_\alpha}$ gives \eqref{eq:main} with the stated spacing threshold.
\end{proof}

The spacing threshold for the power-law estimate \eqref{eq:main} cannot be made independent of $\alpha$, even for fixed $B$ and $G$. Choose antiperiodic boundary conditions in the fourth direction and periodic ones in the other three. Set
$A_N=T\cos(2\pi N x^1)\,dx^4$, where $T=i\pi\,\mathrm{diag}(1,-1,0,0,0)$.
Write $V_N=V(A_N)$. In $S V_N S$ the shifts $h=\pm2\pi N e_1$ give the bound $C\sup_k(|k|\,|k+h|)^{-1}=O(N^{-1})$. Both $|k|$ and $|k+h|$ are at least the free gap, and one of the two is at least $\pi N$. Hence $(V_N S)^2=O(N^{-1})$ in operator norm. Since $V_NS$ is uniformly bounded, the factorization $(1+V_NS)^{-1}=(1-V_NS)[1-(V_NS)^2]^{-1}$ gives a uniform bound on the continuum inverse for large $N$.

At spacing $1/N$, every temporal link is $\exp(T/N)$ and every spatial link is the identity. On the component of $\overline{\mathbf5}$ where $R_*(T)$ has eigenvalue $-i\pi$, the temporal link phase cancels the translation phase of the mode $k=(0,0,0,\pi)$. The overlap operator therefore has an exact zero mode. Yet for $\alpha_N=1/\log(1+2\pi N)$ one has $\|A_N\|_{\mathcal W_{\alpha_N}}=e\|T\|$. A common spacing threshold for these power-law bounds would thus fail to ensure nonvanishing; theorem~\ref{thm:main} is not contradicted, since $B\varepsilon^{\alpha_N}\to\|T\|$ violates its condition $B\varepsilon^\alpha\le\omega_0$. The amplitudes in this example stay at momenta of order $1/\varepsilon$, so their Fourier tails do not vanish uniformly.

On a compact smooth family with a common gap, $\|b\|_{\mathcal W_1}\le C\|b\|_{H^4}\le C'\|b\|_{C^4}$. After the orientation identification in \eqref{eq:orientation}, the imaginary part of Hsu's proposed estimate (equation~(D.7) in \cite{Hsu:2026gmi}) thus holds with $r=4$ in the stated tangent norm, an $O(\varepsilon)$ error, and $\operatorname{Im} C_a=0$. If one keeps Craig's axes fixed for both Weyl operators, this identification amounts to conjugating Hsu's relative phase.

Along a piecewise $C^1$ path $A(t)$ satisfying the same bounds, lift each phase continuously to the real line and denote its change by $\Delta\Phi$. Integration gives
\begin{equation}
 |\Delta\Phi_{\mathrm{ov}}-\Delta\Phi_{\mathrm{CW}}|
 \le C\varepsilon^\alpha\int_0^1\|\dot A(t)\|_{\mathcal W_\alpha}\,dt .
 \label{eq:path}
\end{equation}
This is the integration criterion of proposition~3 in \cite{Hsu:2026gmi}; with the relative normalization fixed at one end of a gapped path, it gives uniform convergence on families of such paths whose lengths have a common bound.

\subsection{Low-order terms in the modulus comparison}
\label{sec:modulus}

The two-, three- and four-point gauge-field amplitudes survive in the logarithmic derivative of the modulus, $\delta_b\log|W|=\operatorname{Re}\,\delta_b\log W$, even though they drop out of the phase for our anomaly-free representation. Let $L_a(A;b)$ be the real part of the terms of total degree at most four in $\Tr[J_a\sum_{\ell=0}^3(-K_a)^\ell P_+]$, counting $b$ as one insertion. For a continuum potential $C$ and its variation $d$, define $L_c(C;d)$ as the real part of
\begin{equation}
 \sum_{\ell=0}^3(-1)^\ell\tr^{Q_0}(K_-^\ell J_-)
  -\tfrac12\Res[T(\log Q-\log Q_0)]
 \label{eq:low}
\end{equation}
with $P,K_-,J_-,Q,T$ evaluated at $C,d$. Counting $d$ as one insertion, \eqref{eq:low} is a polynomial of degree at most four in the external fields. The measure contributes the purely imaginary term $-i\mathfrak L_b$ in \eqref{eq:ov}. The five-point cancellation and the subsequent trace estimates in appendix~\ref{app:odd} apply before taking the real part. Under the hypotheses of theorem~\ref{thm:main}, they give
\begin{equation}
 \left|\operatorname{Re}\,\delta_b\log
    \frac{W_{\mathrm{ov}}^{(a)}[U_a(A)]}{W_{\mathrm{CW}}^{(a)}[U_a(A)]}
          -L_a(A;b)+L_c(A^r;b^r)\right|
  \le C_{B,G}\bigl(\omega_\varepsilon(b)+b_0\omega_\varepsilon(A)\bigr).
 \label{eq:real}
\end{equation}
For $\|A\|_{\mathcal W_1}\le B$, the right-hand side is at most $C_{B,G}\varepsilon\|b\|_{\mathcal W_1}$. The one-point function vanishes because the generators are traceless. This leaves the two-, three- and four-point functions, evaluated at $A$ on the lattice and at $A^r$ in the continuum. Their finite parts can depend nonlocally on the field, as the vacuum polarization illustrates. Because reconstruction is nonlinear, a polynomial in $A^r$ can contain arbitrarily high powers of $A$. Section~\ref{sec:two} determines the quadratic coefficient. Section~\ref{sec:ward} bounds the differences of the cubic and quartic vertices and the errors introduced by reconstruction, keeping track of their dependence on the external momenta.

\section{Gauge kinetic counterterm from vacuum polarization}
\label{sec:two}

The two regulators have the same nonlocal vacuum polarization in the continuum limit. Their quadratic difference is a local gauge kinetic term, whose finite coefficient depends on the spectral regularization prescription and scale. Choose anti-Hermitian generators with $-\tr_{\mathbf5}(t^bt^c)=\delta^{bc}/2$ and $-\tr_R(t^bt^c)=T(R)\delta^{bc}$. On the unit torus define
\begin{equation}
 f=dA=t^b f^b,\qquad I_2(A)=\frac14\int_{T^4} f^b_{\mu\nu}f^b_{\mu\nu}\,d^4x,
 \qquad \bar\mu=\mu L .
 \label{eq:gaugeaction}
\end{equation}
Here $f$ is the linear part of the non-Abelian field strength $F=dA+A\wedge A$, and $T(R)$ is the Dynkin index of $R$ in our trace convention. We absorb the gauge coupling into $A$ and define $I_2$ without the overall Yang-Mills factor $1/g^2$. The mass scale $\mu$ of the zeta-regularized determinant fixes the finite part of the counterterm. In section~\ref{sec:response} we used $\bar\mu=1$.

\begin{samepage}
\begin{theorem}\label{thm:two}
Suppose that the Fourier support of $A$ lies in a fixed finite set of torus momenta. Let $[s^2]$ denote the coefficient of $s^2$ in the Taylor expansion about $s=0$. Then
\begin{align}
&
 [s^2]\log\left|\frac{W_{\mathrm{ov}}^{(a)}[U_a(sA)]}{W_{\mathrm{CW}}^{(a)}[U_a(sA)]}\right|
 = c_R(a,\mu) I_2(A)+O(\varepsilon^2\ell_\varepsilon),
 \nonumber\\[0.35 em]
 &
 c_R(a,\mu) =T(R)\left[\frac{\log(a^2\mu^2)}{24\pi^2}
                         +k_f(1)+\frac{5}{72\pi^2}\right].
 \label{eq:vacmatch}
\end{align}
The error is uniform over bounded sets of fields with this Fourier support. For one multiplet in $\mathbf{10}\oplus\overline{\mathbf5}$, $T(R)=2$. The finite lattice constant $k_f(1)$ is defined in \eqref{eq:axialdef}.
\end{theorem}
\end{samepage}

To define the lattice constant, consider one massless Dirac fermion in the fundamental representation on the infinite lattice, with Wilson kernel mass parameter one. Fix a nonzero momentum $h_*$ along a lattice axis. Let $F(t)$ denote a transverse diagonal component of the vacuum polarization at spacing $t$, divided by $|h_*|^2$, computed with $\log W$ and the generator normalization above. Then
\begin{equation}
   k_f(1)=\lim_{t\downarrow0}\left[F(t)-\frac{\log(t^2|h_*|^2)}{24\pi^2}\right].
   \label{eq:axialdef}
\end{equation}
Appendix~\ref{app:vacuum} proves that this limit exists and is independent of the chosen nonzero axial momentum by comparing $h_*$ and $\lambda h_*$, uniformly for $1\le\lambda\le2$. The result is the finite lattice constant (equations~(3.2)-(3.4) in \cite{Alexandrou:1999wr}). Their table~I gives
$k_f(1)=d_{1,1}(1)+d_{1,2}(1)\simeq-0.01161402+0.01893460=0.00732058$.
The continuum spectral calculation in appendix~\ref{app:proofs} supplies $5/(72\pi^2)$, reproducing the fermion-loop $\overline{\rm MS}$ constant $-5/3$. Thus the finite part of $c_R/T(R)$ is approximately $0.0143568$, consistent with the coupling conversion (relation~(3.6) in \cite{Alexandrou:1999wr}) and with the sign appropriate to $\log W$. The counterterm is defined by the limit, not by the quoted rounded value.

With the sign convention of \eqref{eq:vacmatch}, the subtracted ratio is $e^{-C_a}W_{\mathrm{ov}}/W_{\mathrm{CW}}$ when the quadratic part of $C_a$ is $+c_R I_2$. Polarization of the quadratic identity gives the two-point vertex for independent external fields.

The proof, in appendix~\ref{app:proofs}, writes the Weyl log modulus as $\tfrac14\Tr\log(D^\dagger D)$, computes the continuum polarization at the spectral scale $\mu$, and takes the lattice constant from \cite{Alexandrou:1999wr}; appendix~\ref{app:vacuum} bounds the error.

\section{Non-Abelian Ward identities and determinant matching}
\label{sec:ward}

The two-point coefficient must also match the cubic and quartic parts of the Yang-Mills action. Gauge invariance alone does not establish this: it allows nonlocal transverse amplitudes, including torus holonomy dependence. We first bound the nonlocal remainders, then use the non-Abelian Ward identities to identify the remaining local terms. The resulting counterterm is isotropic even when the spin structure distinguishes the axes.

With the normalization of section~\ref{sec:two}, the Yang-Mills functional and its cubic and quartic terms are
\begin{align}
 I(A)&=-\tfrac12\int\tr_{\mathbf5} F_{\mu\nu}F_{\mu\nu}
       =I_2(A)+I_3(A)+I_4(A),\qquad C_{\mu\nu}=[A_\mu,A_\nu],\nonumber\\
 I_3(A)&=-\int\tr_{\mathbf5} f_{\mu\nu}C_{\mu\nu},\qquad
 I_4(A)=-\tfrac12\int\tr_{\mathbf5} C_{\mu\nu}C_{\mu\nu}.
 \label{eq:kinetic}
\end{align}
Both spacetime indices in \eqref{eq:kinetic} are summed. Write $\mathcal M_a(A)=\log|W_{\mathrm{ov}}^{(a)}[U_a(A)]|$ and $\mathcal M(A)=\log|W_{\rm cont}(A)|$ for the logarithms of the lattice and continuum moduli. On the same links, their difference is $\mathcal M_a(A)-\mathcal M(A^r(A))$.

\begin{theorem}\label{thm:ward}
For $n=3,4$ and smooth $A$ in a bounded $\mathcal W_3$ ball,
\begin{equation}
 [s^n]\big\{\mathcal M_a(sA)-\mathcal M(A^r(sA))\big\}
     =c_R(a,\mu) I_n(A)+O(\varepsilon).
 \label{eq:wardmatch}
\end{equation}
The coefficient, including its finite part, is the one in \eqref{eq:vacmatch}. The error is uniform on the ball. Polarization yields the corresponding estimate for mixed derivatives in independent external fields.
\end{theorem}

The proof is in appendix~\ref{app:proofs}. Taylor subtraction in the external momenta isolates, at each order, a local polynomial of degree at most $4-n$ and a remainder that vanishes with the spacing; the Ward identity of a background gauge field then fixes the polynomials at three and four points in terms of the two-point coefficient.

\begin{corollary}\label{cor:isotropy}
Let the spin structure be antiperiodic in at least one direction and otherwise arbitrary. Let $C_a=\int_{T^4}c_a^{\mu\nu\rho\sigma}\tr_{\mathbf5}F_{\mu\nu}F_{\rho\sigma}$ be a gauge-invariant local functional of dimension four, with the coefficient tensor antisymmetric within each index pair and symmetric between the pairs. Suppose that $\log|W_{\mathrm{ov}}^{(a)}/W_{\mathrm{CW}}^{(a)}|-C_a$ has a vanishing quadratic Taylor coefficient as $a\to0$ on fields with fixed finite Fourier support. Then $c_a^{\mu\nu\rho\sigma}$ equals the isotropic tensor of $c_R(a,\mu)I$ up to $o(1)$ and a totally antisymmetric part, which multiplies the topological density and has no variation on the trivial bundle.
\end{corollary}
\begin{proof}
By theorem~\ref{thm:two}, the quadratic coefficient of $C_a-c_RI$ tends to zero on every fixed Fourier support. Its symbol is $c_a^{\mu\nu\rho\sigma}$ minus the isotropic tensor, contracted with the linearized curvatures; letting the support range over all momentum pairs $\pm h$ and polarizations determines the tensor up to its totally antisymmetric part, which drops out of the quadratic form. The isotropy itself comes from lemma~\ref{lem:evenbubble}: the local part of the two-point difference is fixed in infinite volume, where hypercubic averaging leaves only the transverse tensor, and the winding corrections of the two regularizations agree to $O(\varepsilon^2)$ for every spin structure. By gauge invariance the cubic and quartic parts of $C_a$ are determined by $c_a^{\mu\nu\rho\sigma}$, in agreement with theorem~\ref{thm:ward}.
\end{proof}

A constant noncommuting potential isolates the quartic part of the action and tests its normalization. Embed $t_j=-i\sigma_j/2$ in a two-dimensional block of $SU(5)$, and take $A_1=u t_1$, $A_2=v t_2$, with the other components zero. Then $F_{12}=uvt_3$, so $I(A)=u^2v^2/2$ on the unit torus. Polarizing \eqref{eq:wardmatch} gives $c_R/2+O(\varepsilon)$ for the coefficient of $u^2v^2$ in the difference of the log moduli. For commuting constant potentials, $I=0$, although each determinant can still depend on the torus holonomies, as the example in appendix~\ref{app:vacuum} illustrates.

\subsection{Matching at finite background amplitude}

The gap and the weak-field condition play different roles in this comparison. To see this, return to the noncommuting constant potentials above, with $u=v=s$. Take antiperiodic boundary conditions in time and periodic ones in space. For any time-independent spatial potential with $A_4=0$, the spatial Dirac operator $D_{\rm sp}=\sum_{j=1}^3\gamma_j\nabla_j$ anticommutes with $\gamma_4$ and commutes with $\partial_4$. Hence
\begin{align}
 D_A^\dagger D_A=-\partial_4^2+D_{\rm sp}^\dagger D_{\rm sp}\ \ge\ \pi^2
\end{align}
in our dimensionless units. For the constant example, the mode $k=(0,0,0,\pi)$ is invariant under $V(A)S$. On this mode, $S=-i\gamma_4/\pi$ anticommutes with $V(A)$; their product is anti-Hermitian and has spectral radius $\|V(A)\|/\pi$. This grows linearly in $|s|$. Thus the vacuum Neumann series for the inverse fails at sufficiently large $|s|$, even though the field strength is nonzero and the gap remains open. The following estimate still applies, uniformly on every bounded interval of $s$.

\begin{theorem}\label{thm:finite}
Use the reconstruction of section~\ref{sec:result} and fix the torus, its spin structure, the representation and the spectral scale. For smooth fields satisfying $\|A\|_{\mathcal W_3}\le B$ and $\|D_A^{-1}\|\le G$, the two weights are nonzero for all sufficiently small $\varepsilon$, uniformly over this class. With the orientation in \eqref{eq:orientation}, every such field and every smooth variation $b$ satisfy
\begin{equation}
 \left|\delta_b\log\frac{W_{\mathrm{ov}}^{(a)}[U_a(A)]}{W_{\mathrm{CW}}^{(a)}[U_a(A)]}
            -c_R(a,\mu)\,\delta_b I(A)\right|
      \le C_{B,G}\varepsilon\,\|b\|_{\mathcal W_3}.
 \label{eq:finitevariation}
\end{equation}
\end{theorem}

\begin{proof}
The terms $L_a(A;b)$ and $L_c(A;b)$ in \eqref{eq:real} are the variations of the fourth-order Taylor polynomials of $\mathcal M_a$ and $\mathcal M$ at zero field. Summing \eqref{eq:alljets} against the Fourier coefficients bounds the difference of these variations at the same $(A,b)$ after the counterterm is subtracted. We must also account for evaluating the continuum polynomial at the reconstructed field and for the two complex remainders beginning at fifth order. Write $\mathcal E_a$ for the subtracted variation inside the absolute value in \eqref{eq:finitevariation}. The three contributions give
\begin{align}
 |\mathcal E_a|\le
 C_B\varepsilon\|b\|_{\mathcal W_3}
 +C_B\varepsilon^2\ell_\varepsilon\|b\|_{\mathcal W_3}
 +C_{B,G}\varepsilon\|b\|_{\mathcal W_1}.
\end{align}
The second term follows from appendix~\ref{app:rec}. For the third, appendix~\ref{app:odd} treats the five-point contact diagrams together with the five-propagator trace, then bounds the six-propagator remainder. Each term is linear in $b$. Since $\varepsilon\ell_\varepsilon$ stays bounded and $\|b\|_{\mathcal W_1}\le\|b\|_{\mathcal W_3}$, these estimates give \eqref{eq:finitevariation}. Changing the free spectral weight to $Q_0/\bar\mu^2$ adds a local residue of degree at most four, which is included in the matching at the scale of \eqref{eq:vacmatch}. The trace-class remainder is independent of that scale.
\end{proof}

Since $\|A\|_{\mathcal W_3}\le C_s\|A\|_{H^s}$ for $s>5$, a uniform $C^6$ bound suffices. With the chirality identification \eqref{eq:orientation}, theorem~\ref{thm:finite} gives the full complex response proposed by Hsu (equation~(D.7) in \cite{Hsu:2026gmi}), with a specified counterterm and an $O(\varepsilon)$ rate. Integrating along gapped paths of uniformly bounded length compares the subtracted ratios after normalization at one endpoint.

The Wilson plaquette action gives a counterterm expressed entirely in the links. Let $U_{\mu\nu}(x)$ be the ordered product of links around the elementary square in the $\mu\nu$ plane at $x$, and set
\begin{equation}
 C_a^{\rm link}(U)=2c_R(a,\mu)\sum_{x,\,\mu<\nu}
                 \operatorname{Re}\tr_{\mathbf5}\big(1-U_{\mu\nu}(x)\big).
 \label{eq:linkct}
\end{equation}
To compare the two counterterms, choose radial gauge about the centre of each square. In the frame obtained by radial parallel transport, the potential at displacement $z$ is the integral of $t z^\nu F_{\nu\mu}(tz)$ over $0\le t\le1$. The term linear in $z$ integrates around the square to $\pm\varepsilon^2 F_{\mu\nu}(0)$. The term quadratic in $z$ integrates to zero by inversion about the centre. The integral of the cubic remainder is $O_B(\varepsilon^4)$, and products of two edge integrals start at the same order. Thus the real trace of one minus the plaquette holonomy equals $-\varepsilon^4\tr_{\mathbf5} F_{\mu\nu}(0)^2/2+O_B(\varepsilon^6)$. These estimates and the $O_B(\varepsilon^2)$ midpoint quadrature error for the action density require at most three derivatives of the gauge potential. Differentiating in the direction $b$ gives the same bounds multiplied by $\|b\|_{\mathcal W_3}$. Consequently, using \eqref{eq:linkct} in \eqref{eq:finitevariation} changes the left-hand side by at most $C_B\varepsilon^2\ell_\varepsilon\|b\|_{\mathcal W_3}$.

\subsection{Growing external momentum cutoffs}
How far the external momenta may grow is governed by the vacuum polarization. Suppose that the sum of the Fourier amplitudes is bounded while the external momentum cutoff $M/L$ increases as $a$ decreases. For the phase, \eqref{eq:phasefine} already gives uniform convergence when $\varepsilon M\to0$, since $\omega_\varepsilon(f)\le\varepsilon(1+M)\|f\|_{\mathcal W_0}$. For the full complex weight, the bubble and contact diagrams of the vacuum polarization are variations of $D^\dagger D$, whose evenness under reversal of the lattice step eliminates the term linear in the spacing before integration over the loop momentum.

\begin{corollary}\label{cor:band}
Suppose that $A$ and $b$ have Fourier support in $|h|\le M$, with $M\ge1$, $\|A\|_{\mathcal W_0}\le B$ and $\|D_A^{-1}\|\le G$. Both weights are nonzero when $\varepsilon M$ is sufficiently small, with a threshold depending on $B,G$. For either $C_a=c_R I$ or $C_a=C_a^{\rm link}$ (evaluated along $U_a(A)$ in the latter case),
\begin{equation}
 \left|\delta_b\log(W_{\mathrm{ov}}^{(a)}/W_{\mathrm{CW}}^{(a)})-\delta_b C_a\right|
 \le C_{B,G}\big(\varepsilon M^2+\varepsilon^2\ell_\varepsilon M^4\big)\|b\|_{\mathcal W_0}.
 \label{eq:growing}
\end{equation}
In particular, for either subtraction the residual variation tends uniformly to zero on the subclass with $\|b\|_{\mathcal W_0}\le1$ if
\begin{equation}
 M=o(M_*(\varepsilon)),\qquad M_*(\varepsilon)=\varepsilon^{-1/2}\ell_\varepsilon^{-1/4}.
 \label{eq:bandscale}
\end{equation}
\end{corollary}
The proof is given in appendix~\ref{app:proofs}.

A single transverse Abelian mode shows that the window \eqref{eq:bandscale} is necessary for the simultaneous use of the two subtractions (appendix~\ref{app:proofs}); the optimal cutoff for either prescription separately remains open.

\section{Reconstruction with flat collars}
\label{sec:collar}

Hsu's collar reconstruction preserves the link transports but concentrates curvature inside the cells. It supplies the product regions used in his proof of unnormalized site-reflection positivity at zero theta angle and zero Yukawa coupling; a normalized state and Hamiltonian reconstruction require further results (app.\,F.5 in \cite{Hsu:2026gmi}). We compare its phase and quadratic log modulus with the overlap prescription. The phase variations agree, whereas the gauge kinetic counterterm depends on the profile within each cell.

Fix $0<\delta<1/2$. With $r(v)=e^{-1/v}$ for $v>0$ and zero otherwise, set
\begin{equation}
 s(v)=\frac{r(v)}{r(v)+r(1-v)},\qquad
 \vartheta(u)=\frac\pi2s\Big(\frac{u-\delta}{1-2\delta}\Big),\qquad
 \eta_\delta(v)=\cos\vartheta(|v|)\quad(|v|\le1),
 \label{eq:collarstep}
\end{equation}
and $\eta_\delta(v)=0$ outside this interval. Replace Craig's copying weights by $\alpha_i(x)=\prod_\nu\eta_\delta(x_\nu/\varepsilon-i_\nu)$, so that $\rho_i=\alpha_i^2$ and $\sum_i\rho_i=1$. Within distance $\delta\varepsilon$ of each lattice plane, the reconstructed connection has vanishing normal component and tangential components independent of the normal coordinate. The construction reproduces the input links exactly (proposition~7 in \cite{Hsu:2026gmi}).

The first moment of these weights is $\sum_j\rho_jx_j^\mu=\varepsilon\sin^2\vartheta(x_\mu/\varepsilon)$ in the cell $[0,\varepsilon]^4$. Its derivative defines the periodic profile
\begin{equation}
 p(y)=\frac{d}{dy}\sin^2\vartheta(y),\qquad
 \int_0^1p(y)\,dy=1 .
 \label{eq:collarprofile}
\end{equation}
Writing $R^{\rm c}=dA^r|_{A=0}$ for the linearized reconstruction, appendix~\ref{app:collar} proves
\begin{equation}
 [d(R^{\rm c}u)]_{\mu\nu}(x)
 =p(x_\mu/\varepsilon)p(x_\nu/\varepsilon)(du)_{\mu\nu}(x)+O_u(\varepsilon)
 \qquad(\text{no sum}).
 \label{eq:collarcurv}
\end{equation}
Thus the quadratic gauge action is multiplied in the limit by a profile constant. The quadratic determinant also resolves the Fourier components of the profile, giving a second constant:
\begin{align}
 \kappa(\delta)&=\Big(\int_0^1p^2\Big)^2
 =\Big[\frac{\pi^2\mathcal J}{4(1-2\delta)}\Big]^2,
 &\mathcal J&=\int_0^1s'(t)^2\sin^2(\pi s(t))\,dt,
 \label{eq:kappaXi}\\
 \Xi(\delta)&=\sum_{n\in\mathbb Z^2\setminus\{0\}}
 |\hat p_{n_1}\hat p_{n_2}|^2\log(4\pi^2|n|^2),
 &\hat p_n&=\int_0^1p(y)e^{-2\pi iny}\,dy .
 \label{eq:Xi}
\end{align}
Since $p$ is supported in an interval of length $1-2\delta$, Cauchy-Schwarz gives $\kappa\ge(1-2\delta)^{-2}$. Equality requires a constant profile between the collars and can only be approached by smooth profiles. For Hsu's particular step,
\begin{equation}
 \mathcal J\simeq0.9515011,\qquad
 \kappa(\delta)\simeq\Big[\frac{2.347735}{1-2\delta}\Big]^2 .
 \label{eq:kappanum}
\end{equation}
For this step, the limiting value of $\kappa$ as $\delta\to0$ is approximately $5.51$.

\begin{table}[t]
\centering
\begin{tabular}{ccccc}
\hline
$\delta$ & $\to0$ & 0.05 & 0.15 & 0.25\\
\hline
$\kappa(\delta)$ & 5.51 & 6.80 & 11.25 & 22.05\\
$\Xi(\delta)$ & 18.8 & 25.0 & 47.7 & 109.4\\
\hline
\end{tabular}
\caption{Profile constants for \eqref{eq:collarstep}, evaluated by periodic quadrature and Fourier summation.}
\label{tab:collar}
\end{table}

\begin{samepage}
Let $W_{\rm c}^{(a)}[U]$ be the continuum weight \eqref{eq:cwdef} evaluated on the collar reconstruction.
\begin{theorem}\label{thm:collar}
Fix the torus, spin structure, representation, spectral scale, $\delta\in(0,1/2)$ and bounds $B,G$.
\begin{enumerate}
\item[(a)] For smooth fields with $\|A\|_{\mathcal W_1}\le B$ and $\|D_A^{-1}\|\le G$, both weights are nonzero for sufficiently small $\varepsilon$, and
\begin{equation}
 \Big|\operatorname{Im}\,\delta_b\log\frac{W_{\mathrm{ov}}^{(a)}[U_a(A)]}{W_{\rm c}^{(a)}[U_a(A)]}\Big|
 \le C_{B,G,\delta}\varepsilon\|b\|_{\mathcal W_1}
 \label{eq:collarphase}
\end{equation}
for every smooth variation $b$.
\item[(b)] On bounded sets of fields with a fixed finite Fourier support,
\begin{equation}
 [s^2]\log\left|\frac{W_{\mathrm{ov}}^{(a)}[U_a(sA)]}{W_{\rm c}^{(a)}[U_a(sA)]}\right|
 =c_R^{\rm c}(a,\mu)I_2(A)+O(\varepsilon\ell_\varepsilon),
 \label{eq:collarmatch}
\end{equation}
where
\begin{equation}
 c_R^{\rm c}(a,\mu)=T(R)\Big[
 \kappa\frac{\log(a^2\mu^2)}{24\pi^2}
 +k_f(1)+\kappa\frac{5}{72\pi^2}-\frac{\Xi}{24\pi^2}\Big].
 \label{eq:collarcR}
\end{equation}
\end{enumerate}
\end{theorem}
\end{samepage}

The proof is in appendix~\ref{app:collar}. Setting $\kappa=1$ and $\Xi=0$ recovers the coefficient for Craig's interpolation. The factor $\kappa$ reflects the quadratic curvature energy at the cell scale; $\Xi$ records its distribution among Fourier modes. Their contribution is local and can be absorbed into the gauge coupling. Comparisons between reconstructions must therefore hold renormalized couplings fixed. Only the quadratic coefficient is matched here; extending the modulus comparison to finite backgrounds also requires the three- and four-point remainders.

\section{Numerical convergence and logarithmic lattice artifacts}
\label{sec:numerics}

The uniform estimates are upper bounds. Here we test the logarithmic variations at the original background and compute the leading logarithmic lattice artifact in the two-point function.

\subsection{Numerical comparison in a hypercharge background}

We restrict ourselves to the hypercharge direction of one generation: the charges $q=6Y\in\{1,-4,2,-3,6\}$, with multiplicities $m_q=\{6,3,3,2,1\}$, obey $\sum m_qq=\sum m_qq^3=0$, $\sum m_qq^5=4320$ and $\sum m_qq^2=120$. The background and variation depend only on $x^1$:
\begin{align}
&
\begin{aligned}
 B_1&=0.37\beta,\qquad B_2=\beta[0.9\cos(2\pi x^1+0.3)+0.4\sin(4\pi x^1-0.5)],\\
 B_3&=\beta[0.8\sin(2\pi x^1+1.1)+0.3\cos(4\pi x^1+0.2)],\\
 B_4&=\beta[0.7\cos(2\pi x^1-0.7)+0.5\sin(2\pi x^1+0.4)],
\end{aligned}
\\[0.5 em]
&
\begin{aligned}
 b_1&=0.21,\qquad b_2=0.6\sin(2\pi x^1-0.2)+0.2\cos(4\pi x^1+0.9),\\
 b_3&=0.5\cos(2\pi x^1+0.6)+0.25\sin(4\pi x^1-0.1),\\
 b_4&=0.4\sin(2\pi x^1+0.8)+0.3\cos(2\pi x^1-1.3).
\end{aligned}
\end{align}
Fermions are antiperiodic in $x^4$ and periodic in the other directions. Each component is constant along its corresponding link, so the exact links are $\exp(iq\varepsilon B_\mu)$. At each transverse momentum $(k_2,k_3,k_4)$, the overlap operator is a $4N\times4N$ matrix. Diagonalizing the Hermitian Wilson kernel and differentiating its sign function by divided differences~\cite{Daleckii:1965} gives
\begin{align}
 \Phi_N=\sum_qm_q\sum_k\Tr[(\delta_bD_a)D_a^{-1}P_+] .
\end{align}
Its imaginary part is the dynamical phase variation; its real part is the log-modulus variation. This computation omits reconstruction and the measure term, which is $O(\varepsilon^6)$ on this fixed smooth family.

All computations use double precision in \texttt{NumPy}, with the momentum sums distributed over independent single-threaded processes. The lattices with $N\le 64$ , the continuum extrapolation and the linear response were computed on a 24-core Ryzen workstation, where $N=64$  took about an hour per background. The two $N=80$  lattices took one to two hours each on 100-112 cores of a cluster node with two 64-core EPYC 7J13 processors.

The continuum phase response is evaluated with Fourier cutoff $|k|\le2\pi\Lambda$, $\Lambda\le24$. Contributions with at most four external insertions vanish, counting the variation as one. This cancellation and the odd leading ultraviolet tail motivate extrapolation in $\Lambda^{-2}$; the spread obtained by also using $\Lambda^{-3}$ estimates the extrapolation uncertainty. The resulting value is denoted by $\Phi_\infty$. At $\Lambda=16$, $\beta=0.75$, the vacuum partial sums through five, seven, nine, eleven and thirteen points are $0.28,-0.37,-0.003,-0.88,1.23$, while the resolvent value is $1.2398$. These finite-matrix values illustrate the importance of retaining the resolvent at the larger amplitude.

Table~\ref{tab:numerics} compares the imaginary parts and lists
\begin{align}
 \Delta_N=\operatorname{Im}\Phi_N-\Phi_\infty,\qquad
 \mathcal R_N=\operatorname{Re}\Phi_N-\frac{120\log(\varepsilon^2)}{24\pi^2}\,\delta_b I_B,
 \qquad I_B=\tfrac14\int F_B^2 .
\end{align}
Only the logarithmic counterterm is subtracted in $\mathcal R_N$. Its limiting constant is fitted, so this part of the computation does not test the finite counterterm or the continuum Quillen normalization.

\begin{table}[t]
\centering
\small
\begin{tabular}{r|ccc|ccc}
\hline
 & \multicolumn{3}{c|}{$\beta=0.25$} & \multicolumn{3}{c}{$\beta=0.75$}\\
$N$ & $\Delta_N$ & $N^2\Delta_N$ & $\mathcal R_N$ & $\Delta_N$ & $N^2\Delta_N$ & $\mathcal R_N$\\
\hline
16 & $-1.999\times10^{-4}$ & $-0.051180$ & $1.56205$ & $1.271\times10^{-1}$ & $32.544$ & $2.58413$\\
24 & $-6.895\times10^{-5}$ & $-0.039717$ & $1.55360$ & $5.642\times10^{-2}$ & $32.496$ & $2.33431$\\
32 & $-3.506\times10^{-5}$ & $-0.035906$ & $1.54471$ & $3.169\times10^{-2}$ & $32.447$ & $2.24008$\\
40 & $-2.138\times10^{-5}$ & $-0.034213$ & $1.53979$ & $2.026\times10^{-2}$ & $32.418$ & $2.19630$\\
48 & $-1.446\times10^{-5}$ & $-0.033325$ & $1.53701$ & $1.406\times10^{-2}$ & $32.399$ & $2.17283$\\
64 & $-7.932\times10^{-6}$ & $-0.032491$ & $1.53426$ & $7.904\times10^{-3}$ & $32.375$ & $2.15003$\\
80 & $-5.024\times10^{-6}$ & $-0.032157$ & $1.53305$ & $5.056\times10^{-3}$ & $32.360$ & $2.13984$\\
\hline
\end{tabular}
\caption{Phase-response difference $\Delta_N$ and log-modulus response $\mathcal R_N$ after subtraction of the logarithmic counterterm. The continuum phase response $\Phi_\infty$ is extrapolated from Fourier cutoffs up to $\Lambda=24$. The real part is not compared with a separately computed continuum value.}
\label{tab:numerics}
\end{table}

\begin{figure}[t]
\centering
\includegraphics{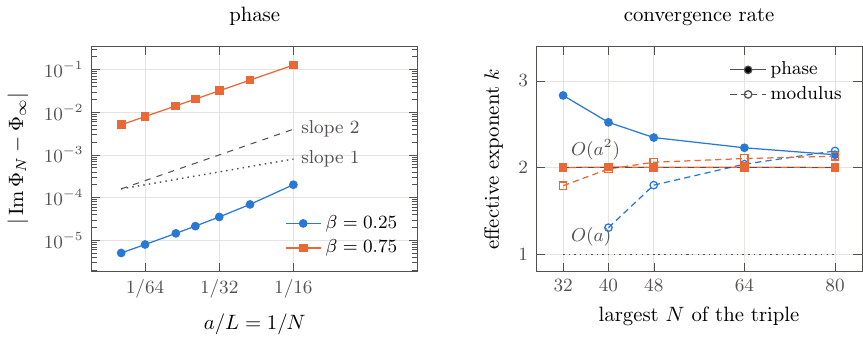}
\caption{Logarithmic variations at the original hypercharge background. Left: absolute difference of the lattice and extrapolated continuum phase responses, with slope-one and slope-two guides. Right: effective exponents from consecutive triples of lattices, for the phase response (filled markers) and the log-modulus response after logarithmic subtraction (open markers). Colours label the two amplitudes as on the left. Reconstruction and the measure term are omitted.}
\label{fig:convergence}
\end{figure}

The phase data are consistent with an $O(\varepsilon^2)$ error. For $\beta=0.75$, $N^2\Delta_N$ varies by $0.6\%$ over $16\le N\le80$. Fitting $X_0+p_2\varepsilon^2+p_4\varepsilon^4$ to $\operatorname{Im}\Phi_N$ on $24\le N\le80$ gives $X_0=1.239895$, within $8.0\times10^{-6}$ of $\Phi_\infty=1.239903$, whose estimated uncertainty is $1.2\times10^{-5}$. At $\beta=0.25$, the smaller quadratic coefficient makes higher powers more visible. Figure~\ref{fig:convergence} also uses the effective exponent $k$, defined for three spacings by
\begin{align}
 \frac{\varepsilon_1^k-\varepsilon_2^k}{\varepsilon_2^k-\varepsilon_3^k}
 =\frac{X_1-X_2}{X_2-X_3}.
\end{align}
This diagnostic eliminates an unknown additive limit. The exponents near two support the observed phase rate; they do not establish a uniform asymptotic expansion.

For the real part, adding $\varepsilon^2\log\varepsilon$ improves a simple extrapolation test. When we fit $N<80$ and predict $N=80$, the forms $X_0+p_2\varepsilon^2+p_4\varepsilon^4+p_6\varepsilon^6$ and $X_0+(p_2+p_2'\log\varepsilon)\varepsilon^2+p_4\varepsilon^4$ give errors $5.4\times10^{-4}$ and $4.1\times10^{-5}$, respectively, at $\beta=0.75$. Effective exponents above two can arise from cancellations at these spacings; they alone do not identify a logarithm. The two-point calculation below determines a nonzero logarithmic coefficient analytically.

\subsection{Logarithmic correction to vacuum polarization}

In the free overlap vacuum polarization, a transverse harmonic contributes
\begin{align}
 c_p\big[\log(\varepsilon^2p^2)+\mathrm{const}
       +u_6\varepsilon^2p^2\log(\varepsilon^2p^2)+O(\varepsilon^2p^2)\big]
\end{align}
to the part of $\operatorname{Re}\Phi_N$ linear in the background. In the hypercharge normalization above, $c_p=120[\delta_b I_B]_p/(24\pi^2)$, where the brackets denote the contribution of that harmonic.

The coefficient $u_6$ follows from the soft-momentum expansion of $D_\chi=D(1-\varepsilon D/2)^{-1}$, where $D$ denotes the overlap operator in dimensionless units. This expression is used only away from the eigenvalue $2/\varepsilon$: near doubler momenta, the separate terms in $\log\det D_\chi-\log\det(1+\varepsilon D_\chi/2)$ are singular, although their sum is analytic in the external momentum. In the continuum expansion below, $D_\mu=\partial_\mu+R_*(A_\mu)$ is the covariant derivative and $D^2=\sum_\mu D_\mu^2$. For exact link transporters, expanding the polar factor of the Wilson kernel gives
\begin{equation}
 D_\chi=\gamma\cdot D+\varepsilon^2\Big[\tfrac16\sum_\mu\gamma_\mu D_\mu^3
 +\tfrac14(\gamma\cdot D)^3-\tfrac14\{\gamma\cdot D,D^2\}\Big]+O(\varepsilon^4).
 \label{eq:dchi}
\end{equation}
At first order in $\varepsilon^2$, the last two insertions in $\Tr\log D_\chi$ reduce, after combining bubble and contact terms, to $\tfrac14\Tr(\gamma\cdot D)^2-\tfrac12\Tr D^2$. These local differential traces, and the soft expansion of $\log\det(1+\varepsilon D_\chi/2)$, contribute no external-momentum logarithm. Only the cubic covariant derivative needs to be retained.

The cubic-derivative bubble has an explicit spin trace. For loop momentum $k$, external momentum $p$ and polarization $e$, define, to order $\varepsilon^2$,
\begin{align}
 \widetilde k_\mu=k_\mu-\tfrac{\varepsilon^2}{6}k_\mu^3,\qquad
 v_\mu=e_\mu\big[1-\tfrac{\varepsilon^2}{6}(3k_\mu^2+3k_\mu p_\mu+p_\mu^2)\big].
\end{align}
Writing $\widetilde q=\widetilde{k+p}$ and omitting common colour and normalization factors, the parity-even bubble integrand is
\begin{align}
 -4\,\frac{2(\widetilde k\cdot v)(\widetilde q\cdot v)
 -(\widetilde k\cdot\widetilde q)v^2}{\widetilde k^2\widetilde q^2}.
\end{align}
The two-field contact vertex of the cubic derivative is independent of $p$: the six ordered placements sum to $6k_\mu$. It cannot contribute to the $\varepsilon^2p^4/|k|^4$ term. Expanding the bubble in the region $|p|\ll|k|\ll1/\varepsilon$, the terms responsible for the logarithm are $(C_0+\varepsilon^2C_2)/|k|^4$. For $p\cdot e=0$, angular averaging over $n=k/|k|$ with
\begin{align}
 \left\langle\prod_{\mu=1}^4 n_\mu^{2r_\mu}\right\rangle_{S^3}
 =\frac{\prod_\mu(2r_\mu-1)!!}{4\cdot6\cdots(4+2\sum_\mu r_\mu-2)}
\end{align}
gives
\begin{align}
 C_0=-\tfrac43p^2e^2,\qquad
 C_2=-\tfrac15(p^2)^2e^2+\tfrac{11}{45}
 \left(e^2\sum_\mu p_\mu^4+p^2\sum_\mu p_\mu^2e_\mu^2\right).
\end{align}
The ratio $C_2/(C_0p^2)$ therefore yields, for unit $\hat p,\hat e$,
\begin{equation}
 u_6(\hat p,\hat e)=\frac3{20}-\frac{11}{60}
 \sum_\mu\big(\hat p_\mu^4+\hat p_\mu^2\hat e_\mu^2\big).
 \label{eq:u6general}
\end{equation}
It is $-1/30$ along an axis and $7/120$ for $\hat p\propto(1,1,0,0)$, $\hat e\propto(0,0,1,0)$. This is a specific coefficient of the logarithmic lattice artifacts discussed more generally in~\cite{Symanzik:1983dc,Husung:2019ytz,Husung:2022kvi}. In the axial normalization \eqref{eq:axialdef},
\begin{equation}
 F(t)=k_f(1)+\frac1{24\pi^2}
 \Big[\log(t^2|h_*|^2)-\frac1{30}t^2|h_*|^2\log(t^2|h_*|^2)\Big]+O(t^2).
 \label{eq:axialsix}
\end{equation}
Thus the logarithmic factor in lemma~\ref{lem:evenbubble} cannot in general be removed.

The linear response $\partial_\beta\operatorname{Re}\Phi_N|_{\beta=0}$, computed from the first two variations of the free sign function, provides an independent numerical check. Fits through $N=128$ give $u_6=-0.03334\pm0.00020$ at $p=2\pi$ and $-0.03334\pm0.00013$ at $p=4\pi$; the uncertainties are the spreads over fit ranges and orders. For the two harmonics of the stated background, the coefficients including the factor $\beta$ are $c_{2\pi}=7.18461\beta$, $c_{4\pi}=-4.04000\beta$, giving $r_2'=-\sum_pc_pp^2/15=23.6223\beta$ in $(r_2+r_2'\log\varepsilon)\varepsilon^2$. This determines the logarithmic coefficient at two-point order. It does not determine the additional contribution from reconstruction or establish an error expansion valid to all orders in the background.

\section{Conclusions and discussion}
\label{sec:discussion}

The matching theorems compare the complex logarithmic variations of the two fermion weights on uniformly gapped backgrounds, including fields for which the vacuum expansion fails. Their phase content is nontrivial: anomaly cancellation does not make a chiral determinant real \cite{Ball:1988xg}. For hypercharge normalized by $q=6Y$, one generation has $\sum q^3=0$ but $\sum q^5=4320$. A separated-point five-current response therefore tests the phase without a local-counterterm ambiguity. On $\mathbb R^4$, take insertion points $0,e_1,e_2,e_3,e_4$ and current components $1,1,1,1,2$. With $S(x)=\gamma\cdot x/(2\pi^2|x|^4)$, the 24 oriented cycles have a common denominator $64(2\pi^2)^5$. Their $\gamma_5$ traces are sixteen values of $+4$ and eight of $-4$; the traces without $\gamma_5$ cancel. Including $P_-$ and $i^5$, the fifth derivative of the phase is
\begin{align}
 -\frac{4320}{2(2\pi^2)^5}\,
       \frac{16\cdot4-8\cdot4}{64}
 =-\frac{1080}{(2\pi^2)^5}.
\end{align}
Hsu's Weyl block in the same axes reverses this sign; the pullback in \eqref{eq:orientation} reverses it again. On the unit torus, take the insertion points $0,\lambda e_1,\ldots,\lambda e_4$ with $\lambda\downarrow0$. The displayed coefficient multiplies the leading $\lambda^{-15}$ term. Smearing the insertions in disjoint neighbourhoods and choosing sufficiently small source amplitudes preserves the Dirac gap, so theorem~\ref{thm:main} applies.

The improved fixed-order estimates have a different origin. Inversion of the loop momentum removes the zero-external-momentum five-point term, while reversal of the signed lattice step removes first-order discretization terms from complete graphs. The Ginsparg-Wilson relation gives the following exact identity for the dynamical phase response.

\begin{lemma}\label{lem:gwcontact}
Where $D_a$ is invertible, for real variations,
\begin{equation}
 \operatorname{Im}\Tr[(\delta_bD_a)D_a^{-1}P_+]
 =\operatorname{Im}\Tr[(\delta_bD_a)_{\rm odd}S_a^\chi P_+],\qquad
 S_a^\chi=D_a^{-1}-\varepsilon/2,
 \label{eq:gwcontact}
\end{equation}
where $(\delta_bD_a)_{\rm odd}=[\delta_bD_a-\gamma_5(\delta_bD_a)\gamma_5]/2$.
The operator $S_a^\chi$ is anti-Hermitian and anticommutes with $\gamma_5$. Both factors on the right are even under reversal of the signed step defined in appendix~\ref{app:vacuum}.
\end{lemma}
\begin{proof}
The Ginsparg-Wilson relation and $\gamma_5$-Hermiticity give
\begin{align}
 D_a^{-1}+(D_a^{-1})^\dagger=\varepsilon,\qquad
 \gamma_5D_a^{-1}\gamma_5=(D_a^{-1})^\dagger,
\end{align}
which prove the assertions about $S_a^\chi$.
The contact contribution $(\varepsilon/2)\Tr[(\delta_bD_a)P_+]$ is real by $\gamma_5$-Hermiticity. The chirality-even part of $\delta_bD_a$, multiplied by $S_a^\chi$, is chirality-odd and has zero trace with $P_+$. Finally $D_{-t}=-\gamma_5D_t\gamma_5$ gives the stated evenness.
\end{proof}

This contact subtraction is familiar from the improvement of Ginsparg-Wilson fermions \cite{Niedermayer:1998bi,Chiu:1998gp}. It does not by itself give a uniform second-order bound: loop momenta of order $1/a$ must also be controlled. Proposition~\ref{prop:allorders} proves $O(\varepsilon^2\ell_\varepsilon)$ only for each fixed Taylor coefficient. The numerical results suggest $O(\varepsilon^2)$ for the phase response at the original smooth field, but a bound uniform in the background amplitude remains open. The two-point modulus has a nonzero logarithmic correction, whose coefficient is computed in \eqref{eq:u6general}. Reconstruction can contribute an additional correction of this order; it is not included in the numerical comparison.

For the collar reconstruction we prove phase matching and quadratic modulus matching. Extending the latter to the cubic and quartic amplitudes requires estimates on their nonlocal remainders before the Ward identities can be applied. The growing-momentum window of corollary~\ref{cor:band} is necessary for simultaneous convergence with the continuum and plaquette subtractions; its optimal extent for either subtraction separately is also open.

All results concern prescribed backgrounds without fermion zero modes. Zero modes require comparison of correlators with enough fermion insertions, and Yukawa couplings introduce scalar vertices. A continuum limit of the dynamical gauge theory would further require control of the gauge-field measure.

\acknowledgments
This work was supported in part by the Start-up Research Fund of Southeast University (No.\,RF1028624160).
Large language models (mostly GPT-6 Astra and Kimi K3.1) were used under the author's direction to assist with developing proofs, checking intermediate arguments, and editing the manuscript. The author takes full responsibility for the results and the final manuscript.

\newpage
\appendix
\section{A collection of proofs}
\label{app:proofs}

This appendix collects the longer proofs of sections~\ref{sec:conv}, \ref{sec:two} and \ref{sec:ward}. Lemma~\ref{lem:S} uses the vertex bounds of appendix~\ref{app:vertices}. Theorem~\ref{thm:two} uses the comparison of the vacuum polarizations in appendix~\ref{app:vacuum} and the reconstruction estimates of appendix~\ref{app:rec}. Theorem~\ref{thm:ward} uses lemma~\ref{lem:jetbound} of appendix~\ref{app:localjets} and lemma~\ref{lem:evenbubble}, and corollary~\ref{cor:band} combines these bounds with those of appendices~\ref{app:rec} and \ref{app:odd}.

\begin{proof}[Proof of lemma~\ref{lem:S}]
Let $k$ be an allowed fermion momentum. Throughout the Brillouin zone the free lattice propagator obeys
$\|S_a(k)\|\le C/|k|$. Indeed, the free overlap symbol has a single zero, at which it has the continuum Dirac dispersion, and outside a neighbourhood of that zero the dimensionless inverse is bounded. The antiperiodic boundary condition excludes $k=0$ from the fermion momentum grid. Summing $|k|^{-t}$ over four-dimensional momentum shells gives
\begin{equation}
 \|S_a\|_t+\|S\|_t\le C\big(1+(t-4)^{-1}\big)^{1/t},\qquad t>4 .
 \label{eq:freeSchatten}
\end{equation}
By \eqref{eq:vertexbd}, the nonlinear part of $V_a$ is bounded in operator norm by $C_B\varepsilon\|A\|_\infty^2$. Its variation is bounded by $C_B\varepsilon\|A\|_\infty\|b\|_\infty$. Multiplying each by $S_a$ and applying \eqref{eq:freeSchatten} gives the corresponding Schatten norm bounds.

To estimate the linear term, take an external field with a single Fourier mode of momentum $h$ and unit amplitude. Multiply each linear vertex by its free propagator on the right and let $E_{h,p}$ be the Schatten $p$-norm of the difference of the resulting operators. If $\varepsilon|k|,\varepsilon|h|\le c$ for a sufficiently small fixed $c$, both operators send $k$ to $k+h$ inside the Brillouin zone. Put $u=\varepsilon k$ and $v=\varepsilon h$. The dimensionless free overlap symbol is $1+X/|X|$, with
\begin{align}
 X(u)=i\sum_\mu\gamma^\mu\sin u_\mu+\sum_\mu(1-\cos u_\mu)-1 .
\end{align}
The expansion $1+X/|X|=i\gamma\cdot u+O(|u|^2)$ yields $\|S_a(k)-S(k)\|\le C\varepsilon$. To expand the vertex, differentiate the contour integral for the spectral projector of the Hermitian Wilson operator in appendix~\ref{app:vertices}. The two free resolvents carry momenta $u+v$ and $u$. Between them, the variation of a covariant lattice translation contains the link integral $\varepsilon\int_0^1e^{itv_\mu}dt$. The factor $\varepsilon^{-1}$ in $D_a$ cancels the link length. The first derivatives of the resulting vertex matrix with respect to $u$ and $v$ are uniformly bounded. At $u=v=0$, $H_{W,0}=-\gamma_5$ anticommutes with its link variation, so the first variations of $\operatorname{sign} H_{W,0}$ and $H_{W,0}$ agree. The vertex at the origin is $\gamma^\mu R_*(A_\mu)$ for the chosen polarization. After multiplication by the respective free propagators, the two vertices therefore differ by at most $C\varepsilon(1+|h|/|k|)$ in matrix norm.

On these low-momentum modes the shift is an isometry. Counting the modes shows that the constant term in the last bound contributes at most $C\varepsilon^{1-4/p}$ to the Schatten norm. The second term adds at most $C_p\varepsilon|h|$ by \eqref{eq:freeSchatten}. For $\varepsilon|k|>c$, bound each shifted vertex separately in operator norm and sum the tail of its free propagator. The shell sum of $|k|^{-p}$ above $c/\varepsilon$ is bounded by $(C\varepsilon^{1-4/p})^p/(1-2^{4-p})$. This estimate includes lattice momenta identified modulo a reciprocal lattice vector and continuum momenta outside the zone. For $\varepsilon|h|>c$, \eqref{eq:freeSchatten} bounds each product directly. Combining the cases gives
\begin{equation}
 E_{h,p}\le C_p\big(\varepsilon^{1-4/p}+\min(1,\varepsilon|h|)\big).
 \label{eq:mode}
\end{equation}
After taking $p$-th roots, the constants are bounded uniformly for $p\ge5$. Summing over the external Fourier coefficients and using $\min(1,\varepsilon|h|)\le\varepsilon^\alpha(1+|h|)^\alpha$ establishes \eqref{eq:S5}. We also need the explicit dependence of the operator norm on the field amplitudes. Write $A_s=\|A\|_{\mathcal W_s}$ and $b_s=\|b\|_{\mathcal W_s}$. Taking $p\to\infty$ in \eqref{eq:mode} bounds the error for one linear insertion by $C\varepsilon(1+|h|)$. When $\varepsilon A_0$ is below a fixed constant, \eqref{eq:vertexbd} bounds the nonlinear part of the vertex by $C\varepsilon A_0^2$ and its variation by $C\varepsilon A_0 b_0$. Hence
\begin{equation}
 \|K_a-K\|\le C\varepsilon(A_1+A_0^2),\qquad
 \|J_a-J\|\le C\varepsilon(b_1+A_0 b_0).
 \label{eq:tameop}
\end{equation}
Keeping the minimum in \eqref{eq:mode} gives the more precise operator bounds
\begin{equation}
 \|K_a-K\|\le C_B\omega_\varepsilon(A),\qquad
 \|J_a-J\|\le C_B\omega_\varepsilon(b) \qquad (A_0\le B).
 \label{eq:modop}
\end{equation}
Indeed, the nonlinear errors above are absorbed into the right-hand sides, since $\varepsilon f_0\le\omega_\varepsilon(f)$, where $f_0=\|f\|_{\mathcal W_0}$.
\end{proof}

\begin{proof}[Proof of theorem~\ref{thm:two}]
We first relate the modulus of the Weyl determinant to that of a vector-like Dirac determinant. The overlap operator $D_a$ maps the field-dependent Weyl space onto the fixed output space. Hence $|W_{\mathrm{ov}}|^2=\det_{P_+}(D_aD_a^\dagger)$ up to a field-independent factor. The Ginsparg-Wilson relation and $\gamma_5$-Hermiticity imply that $D_a$ is normal and that $D_a+D_a^\dagger=\varepsilon D_aD_a^\dagger$. On each eigenspace of $D_a^\dagger D_a$ with eigenvalue $0<\lambda<4/\varepsilon^2$, the operator $D_a-D_a^\dagger$ is invertible and exchanges the two fixed chiralities. Their multiplicities are therefore equal. There are no zero modes near the vacuum, so $\Tr\gamma_5=0$ gives equal multiplicities at $\lambda=4/\varepsilon^2$ as well. Thus $\log|W_{\mathrm{ov}}|$ equals one quarter of $\Tr\log(D_a^\dagger D_a)$ up to a constant. The same factor holds in the continuum because, for the Weyl operator $P$, $P^\dagger P$ and $P P^\dagger$ have the same nonzero spectrum. Relative to a Dirac fermion in the fundamental representation, the colour and spin factor for a Weyl multiplet in representation $R$ is $T(R)/(2T(\mathbf5))=T(R)$.

We next compute the continuum polarization on $\mathbb R^4$ at the chosen spectral scale. Take two transverse polarizations $u,v$ in the same colour direction with dimensionless momenta $h,-h$, where $h\ne0$. The second variation of $I_2$ on these fields is $h^2u\cdot v$ per unit volume. Set $Q_D=D_A^\dagger D_A$ on four-component spinors and write $y=x(1-x)$. The second variation of the heat trace $\Tr e^{-tQ_D}$ has three contributions. The covariant Laplacian supplies both a contact term and a pair of vertices linear in $A$. A second pair comes from the spin coupling to $F_{\mu\nu}$. In that order they give
\begin{align}
 \frac{4T(R)\,u\cdot v}{(4\pi t)^2}
 \left[-2t+\int_0^1(2t+t^2h^2)e^{-t h^2 y}\,dx\right].
\end{align}
This expression follows from the second-order heat kernel expansion by completing the square in the loop momentum. The transverse Gaussian moment is $1/(2t)$ per component, and the mixed spin trace vanishes. Integration by parts using $\partial_x[(1-2x)e^{-th^2y}]$ reduces the result to
$T(R)h^2(u\cdot v)\pi^{-2}\int_0^1 y e^{-th^2y}\,dx$.
The endpoint terms at $x=0$ and $x=1$ cancel the contact contribution $-2t$, leaving the factor $h^2$ required by transversality.
The Mellin transform gives $\log|W_{\rm cont}|=-\zeta'_{Q_D/\bar\mu^2}(0)/4$. The coefficient of $h^2u\cdot v$ in its second variation is
\begin{equation}
 \frac{T(R)}{4\pi^2}\int_0^1y\log\frac{h^2 y}{\bar\mu^2}\,dx
 =\frac{T(R)}{24\pi^2}\left(\log\frac{h^2}{\bar\mu^2}-\frac53\right).
 \label{eq:flatquillen}
\end{equation}
Here $\int_0^1y\,dx=1/6$ and $\int_0^1 y\log y\,dx=-5/18$.

For a massless Dirac fermion in the fundamental representation, the overlap calculation (formula~(3.4) in \cite{Alexandrou:1999wr}) gives $\log(\varepsilon^2 h^2)/(24\pi^2)+k_f(1)$ in this normalization, up to terms that vanish with the spacing. The published calculation traces the polarization tensor over its vector indices. Appendix~\ref{app:vacuum} explains how this trace determines the transverse coefficient for our link variables. Multiplying by the factor $T(R)$ and subtracting \eqref{eq:flatquillen} cancels the nonlocal $\log h^2$ term. The remaining coefficient is \eqref{eq:vacmatch} since $\varepsilon\bar\mu=a\mu$.

To bound the error, appendix~\ref{app:vacuum} compares the infinite-volume polarization tensors over a continuous range of momenta on one axis and estimates the finite-volume corrections from loops winding around the torus. Lemma~\ref{lem:evenbubble} bounds the difference after subtraction at all fixed momenta, including $h=0$. It remains to evaluate the continuum determinant on the reconstructed gauge field. Its one-point function vanishes at $A=0$, so only the first derivative of the reconstruction enters the two-point function. Lemma~\ref{lem:foldedrecon} and the ensuing bound on the continuum two-point function give a reconstruction error of $O(\varepsilon^2\ell_\varepsilon)$ for bounded Fourier amplitudes on the fixed support. Summing over the finitely many momentum pairs proves the result.
\end{proof}

\begin{proof}[Proof of theorem~\ref{thm:ward}]
First compare both regulators at the original field $A$. In lemma~\ref{lem:jetbound} the loop momentum runs over the values allowed by the spin boundary conditions. With $n$ external insertions, Taylor subtraction in the external momenta through degree $4-n$ leaves a summand bounded by $C|k|^{-5}$ at large loop momentum $k$. The subtracted lattice and continuum sums have the same limit. The terms added back are polynomials in the external momenta and hence local in position space. Working directly on the torus includes the four-point function at zero external momentum, where the corresponding massless infinite-volume integral would require separate treatment.

Both functionals of $A$ are gauge invariant at nonzero spacing. Each exact link transporter transforms at its two endpoints, and the Quillen metric is invariant under unitary gauge transformations. After subtracting the gauge-invariant functional $c_R I$, write $D_{m,a}=d^m(\mathcal M_a-\mathcal M-c_R I)|_0$ for the derivatives at zero field. At each fixed momentum pair, lemma~\ref{lem:evenbubble} gives $D_{2,a}=O_h(\varepsilon^2\ell_\varepsilon)$ before reconstruction. For any test function $\omega$ with values in the gauge Lie algebra, invariance under the infinitesimal gauge transformation $d\omega+[A,\omega]$ gives
\begin{equation}
 D_{m,a}(d\omega,b_2,\ldots,b_m)
 =-\sum_{j=2}^m D_{m-1,a}(b_2,\ldots,[b_j,\omega],\ldots,b_m).
 \label{eq:nonabelianward}
\end{equation}
This is the Ward identity for a background gauge field. We use it before imposing a Fourier cutoff because the momentum of $[b_j,\omega]$ is the sum of the two input momenta.

At $m=3$ the polynomial part of $D_{3,a}$ has total degree at most one in the independent momenta $p,q$ of its first two legs. By \eqref{eq:nonabelianward}, its contraction with either momentum tends to zero at every fixed tuple of torus momenta. Consider first an exactly transverse tensor polynomial $P(p,q)$ of this degree. In the identity $p_\mu P_{\mu\nu\rho}=0$, the terms linear in $p$ give $P(0,q)=0$. Contraction with $q$ likewise gives $P(p,0)=0$. Every nonzero monomial would then contain both $p$ and $q$, contrary to the degree bound. At $m=4$ the polynomial is a constant tensor. Once the three-point difference tends to zero, so does the right-hand side of \eqref{eq:nonabelianward} at four points. A constant tensor vanishes if contraction in its first index with every momentum gives zero.

To obtain a rate without assuming that the polynomial coefficients stay bounded as $a\to0$, consider the two finite-dimensional spaces of tensor polynomials just described. A polynomial that vanishes on the integer grid vanishes identically. Thus the contractions at all tuples of torus momenta have trivial common kernel in each space. Finitely many of these contractions already define an injective linear map with a fixed left inverse. At the selected momenta the error from the ultraviolet subtraction is $O(\varepsilon)$ by lemma~\ref{lem:jetbound}. Lemma~\ref{lem:evenbubble} starts the induction with the full two-point tensor on the torus for the chosen spin structure. No symmetry of the spin structure under permutations of the axes is needed. Applying the left inverse at three points and then at four points bounds every residual polynomial coefficient by $C\varepsilon$, independently of the external momentum cutoff. Combining these bounds with the Taylor remainders of lemma~\ref{lem:jetbound} gives, for unit polarizations and $H=1+\sum_i|h_i|$,
\begin{align}
&
\begin{aligned}
 |D_{n,a}(h_1,\dots,h_n)| \le C
 \begin{cases}
 \varepsilon^2\ell_\varepsilon H^4,& n=2,\\
 \varepsilon H^2,& n=3,\\
 \varepsilon H(1+\log H),& n=4
 \end{cases}
\quad (\varepsilon H<c),
 \end{aligned} 
 \nonumber\\[0.5 em]
 &
 \begin{aligned}
 |D_{n,a}(h_1,\dots,h_n)| \le C\varepsilon\prod_i(1+|h_i|)^3 \quad (\text{all momenta}).
 \end{aligned} 
 \label{eq:alljets}
\end{align}

In the first line the vertex vanishes unless the external momenta sum to zero. At two points this condition gives $h_2=-h_1$, so the first bound also implies the second in this region since $\varepsilon\ell_\varepsilon$ is bounded. Appendix~\ref{app:localjets} treats large external momenta, including tuples whose sum is a nonzero reciprocal lattice vector. Summing \eqref{eq:alljets} against the absolute values of the Fourier coefficients bounds each multilinear form on $n$ copies of $\mathcal W_3$. Lemma~\ref{lem:foldedrecon} and the multilinear estimates following it bound the change under reconstruction through fourth order by $O_B(\varepsilon^2\ell_\varepsilon)$. At each order the coefficient is obtained by retaining the finitely many terms of that degree in the convergent source expansion. This proves \eqref{eq:wardmatch}.
\end{proof}

\begin{proof}[Proof of corollary~\ref{cor:band}]
With this cutoff, $A_1\le(1+M)B$ and $b_1\le(1+M)b_0$ in the notation of \eqref{eq:tameop}. The difference between the lattice and continuum operators, multiplied by $(1+V(A)S)^{-1}$, thus has norm at most $C_B(1+CBG)\varepsilon M$. When this bound is below $1/2$, the overlap Dirac operator has a uniformly bounded inverse. Applying the same perturbation argument with \eqref{eq:recon} gives invertibility of the continuum operator at $A^r$.

For $\varepsilon M$ this small, no sum of two, three or four external momenta can equal a nonzero reciprocal lattice vector. Summing \eqref{eq:alljets} therefore gives $C_B\varepsilon^2\ell_\varepsilon M^4$ at two points and $C_B\varepsilon M^2$ at three and four points, suppressing the common factor $\|b\|_{\mathcal W_0}$. With the same factor understood, appendix~\ref{app:rec} bounds the change in the variation of the continuum polynomial under reconstruction by $C_B\varepsilon^2\ell_\varepsilon M^4$. For the fifth-order term and the remainder beginning at sixth order, appendix~\ref{app:odd} gives $C_{B,G}\varepsilon M$, including the effect of reconstruction. Only one external field in each difference carries the additional momentum weight.

For the plaquette subtraction, the curvature satisfies $\|\nabla_A^j F\|_\infty\le C_B M^{j+1}$ for $j=0,1,2$. Differentiating in the direction $b$ yields the same powers of $M$ times $\|b\|_{\mathcal W_0}$. The expansion about the plaquette centre and the midpoint quadrature estimate used for \eqref{eq:linkct} then bound the difference of the two action variations by $C_B\varepsilon^2 M^4\|b\|_{\mathcal W_0}$ before multiplication by $c_R$. This proves \eqref{eq:growing}. With $x=M/M_*(\varepsilon)$, the two error factors become $x^2/\sqrt{\ell_\varepsilon}$ and $x^4$, which proves \eqref{eq:bandscale}.
\end{proof}

\paragraph{Sharpness of the window for the two counterterms together.}
A single transverse Abelian mode tests the difference between the continuum and plaquette subtractions as the external momentum grows. Set $\varepsilon=1/N$, $h=2\pi m$ with $0<2m<N$, and $A=tT\cos(hx^1)\,dx^2$, $b=\partial_tA$, $T=-\tfrac{i}{2}\mathrm{diag}(1,-1,0,0,0)$. For fixed small $t\ne0$ the gap is uniform in $m$ and both unweighted Fourier norms stay bounded. With $P_\varepsilon=2\sum_{x,\mu<\nu}\operatorname{Re}\tr_{\mathbf5}(1-U_{\mu\nu}(x))$, only plaquettes in the $x^1x^2$ plane contribute; their holonomies are $\exp[\varepsilon tT(\cos(h(x^1+\varepsilon))-\cos(hx^1))]$, and $I=t^2h^2/4$. The lattice average of the squared difference is $2\sin^2(\varepsilon h/2)$, so
\begin{equation}
 \partial_t(P_\varepsilon-I)
 =\frac{th^2}{2}\left[\left(\frac{\sin(\varepsilon h/2)}{\varepsilon h/2}\right)^2-1\right]+O_t(\varepsilon^4 h^4)
 =-\frac{t\varepsilon^2 h^4}{24}+O_t(\varepsilon^4h^6+\varepsilon^4h^4).
 \label{eq:onemode}
\end{equation}
Fix $B>0$ and $G>\|D_0^{-1}\|$. If $\limsup_{\varepsilon\to0}M/M_*(\varepsilon)>0$, choose $\lambda$ below this limit superior and $h\in2\pi\mathbb N$ with $h\le M$ and $h/M_*(\varepsilon)\to\lambda$ along a subsequence. Then \eqref{eq:vacmatch} and \eqref{eq:onemode} give $c_R\,\partial_t(P_\varepsilon-I)\to tT(R)\lambda^4/(288\pi^2)\ne0$, the difference of the two residual variations, so the two subtractions cannot both converge uniformly on this class: $M=o(M_*)$ is necessary for their simultaneous use. The optimal cutoff for either prescription separately remains open.

\section{Overlap vertices and the measure term}
\label{app:vertices}

Varying the overlap operator in the lattice determinant produces vertices with arbitrarily many external gauge-field insertions. Varying the chiral measure gives a second term involving Craig's measure current. Analyticity in the links bounds the first contribution to all orders. A bound on four derivatives of the current suffices for the second. The constants in both estimates are uniform in the number of lattice sites.

\begin{lemma}\label{lem:vertex}
Let $V_{n,a}(A)$ be the $n$-field vertex, the coefficient of $z^n$ in $D_a(zA)-D_a(0)$. For $n\ge1$ there is a constant $c_0$, independent of both $n$ and the lattice spacing, such that
\begin{align}
 \|V_{n,a}(A)\|&\le c_0^n\varepsilon^{n-1}\|A\|_\infty^n,\nonumber\\
 \|\delta_b V_{n,a}(A)\|&\le n c_0^n\varepsilon^{n-1}\|A\|_\infty^{n-1}\|b\|_\infty .
 \label{eq:vertexbd}
\end{align}
\end{lemma}
\begin{proof}
Consider the gauge potential $zA+wb$ with complex amplitudes $z,w$. Parallel transport along a forward link is a path-ordered exponential. Transport along the reversed link gives its inverse. The analytic continuation of the conjugate fundamental representation sends a link matrix $U$ to $(U^{-1})^{\mathsf T}$. The distance of either oriented link from the identity is at most $\exp(c\varepsilon[|z|\|A\|_\infty+|w|\|b\|_\infty])-1$. For real fields the Hermitian Wilson kernel is a sum of an on-site term and nearest-neighbour covariant translations with fixed spin matrices. The same expression defines its continuation $H_W$ at complex amplitudes. Thus $\|H_W-H_{W,0}\|\le C r$ whenever $\varepsilon(|z|\|A\|_\infty+|w|\|b\|_\infty)<r$ for sufficiently small $r$.

For the free kernel $H_{W,0}$ with Wilson mass parameter one, $v_\mu=1-\cos u_\mu$ gives
\begin{align}
 H_{W,0}(u)^2=\sum_\mu\sin^2u_\mu+\left(\sum_\mu v_\mu-1\right)^2
        =1+2\sum_{\mu<\nu}v_\mu v_\nu\in[1,49].
\end{align}
Choose a fixed contour $\Gamma$ that encloses the positive spectral interval $[1,7]$, excludes the negative interval and stays a positive distance from both intervals. For sufficiently small $r$, expansion about the free resolvent bounds $(\zeta-H_W)^{-1}$ uniformly on $\Gamma$. The Riesz projection $\Pi_+(H_W)=(2\pi i)^{-1}\int_\Gamma(\zeta-H_W)^{-1}d\zeta$ is holomorphic in the sources. With $H_W=\gamma_5 X$ for the Wilson kernel at mass parameter one, the continuation of $D_a=\varepsilon^{-1}[1+\gamma_5(2\Pi_+(H_W)-1)]$ has norm $O(\varepsilon^{-1})$. For real $z,w$, the contour integral projects onto the positive spectral subspace of $H_W$, giving the usual overlap operator. Cauchy's formula on circles of radii $r/(3\varepsilon\|A\|_\infty)$ and $r/(3\varepsilon\|b\|_\infty)$ in the two variables bounds the coefficients of $z^n$ and $z^{n-1}w$, giving \eqref{eq:vertexbd}. The bounds extend by continuity if either field vanishes.
\end{proof}

As in the derivation of the small-field bound (equation~(D.22) in \cite{Hsu:2026gmi}), consider a term in the polynomial part of \eqref{eq:split} with $m\le4$ free propagators and $n\ge5$ external gauge fields. Each of the $m$ vertices contributes one inverse power of $\varepsilon$ and each field insertion one positive power, by \eqref{eq:vertexbd}. The propagator sums satisfy $\|S_a\|_m^m=O(\varepsilon^{m-4})$ for $m<4$, and $\|S_a\|_4^4=O(1+|\log\varepsilon|)$. There are $O(n^4)$ ways to distribute the fields among the ordered vertices and choose the insertion carrying $b$. For $\|A\|_\infty\le B$ and $\varepsilon c_0 B<1/2$, H\"older's inequality for Schatten norms gives a convergent sum over $n\ge5$, bounded by $C_B\varepsilon(1+|\log\varepsilon|)\|b\|_\infty$. This estimate applies to the full complex trace. The terms of degree at most four have vanishing imaginary part by charge conjugation and the colour-trace identities for $R$ in section~\ref{sec:response}.

For the measure term, formula~(34) in \cite{Craig:2026soq} gives $\mathfrak L_b=\sum_I\theta_I j_I(U)$, where $I$ specifies a positively oriented link and a Lie algebra generator, $j_I$ is the measure current, and $\theta_I$ is the corresponding component of $\delta_b U\,U^{-1}$. Write $X_e=\log U_e(A)$. In the fixed global gauge, $\max_e\|X_e\|\le C\varepsilon\|A\|_\infty$ and $\|\delta_b U_e\, U_e^{-1}\|\le C\varepsilon\|b\|_\infty$. Along $U_e(t)=\exp(tX_e)$, $0\le t\le1$, each plaquette matrix differs from the identity by $O(\varepsilon B)$, so the path is admissible for small $\varepsilon$. We use two properties of the current proved in \cite{Craig:2026soq}: it vanishes with its first three link derivatives at $U=1$ (relation~(276) in \cite{Craig:2026soq}), and by the locality estimate (equation~(40) and theorem~4.1 in \cite{Craig:2026soq}) its fourth derivatives have a uniformly bounded sum over ordered derivative indices, independently of the volume. Taylor's formula along the path and the $O(\varepsilon^{-4})$ link count at fixed physical volume give
\begin{equation}
 |\mathfrak L_b|\le C\varepsilon\,\|A\|_\infty^4\|b\|_\infty .
 \label{eq:measure}
\end{equation}
This bound holds for fields of any bounded amplitude. On compact smooth families the exact measure term is $O(\varepsilon^6)$ (app.\,F.3 in \cite{Craig:2026soq} and lemma~4 in \cite{Hsu:2026gmi}). This stronger estimate is used in the fixed-order analysis of appendix~\ref{app:fixedfive} and in the interpretation of the numerical tests of section~\ref{sec:numerics}.

\section{Reconstruction of the gauge field and its variation}
\label{app:rec}

The reconstruction changes both the gauge potential and its variation.
To measure these changes we identify the reconstructed colour bundle with
the trivial bundle in a single periodic frame. We construct this frame
from the vacuum copying map. Hsu uses radial transport in the original
connection (equations~(F.1), (F.7) and~(D.39) in \cite{Hsu:2026gmi}).
Both identifications pull back the same geometric connection; the
estimates below refer to the former. This choice also lets us follow
Fourier modes whose momenta increase as the lattice spacing decreases.

\begin{lemma}\label{lem:reconstruction}
Let $A$ be a smooth gauge potential and $b$ a variation in a periodic frame on the unit torus, with $\|A\|_\infty\le B$. Obtain the link variables by parallel transport along the edges of length $\varepsilon$ and reconstruct the bundle and connection as in app.\,A of \cite{Craig:2026soq}. For sufficiently small $\varepsilon$ the reconstructed colour bundle has a periodic orthonormal frame, smooth in $A$, whose exterior product is the prescribed $SU(5)$ volume element. Write $A^r(A)$ for the gauge potential in this frame and $b^r=\partial_t A^r(A+tb)|_{t=0}$. For $0<\beta\le1$,
\begin{equation}
 \begin{aligned}
 \|A^r-A\|_\infty
 &\le C_B\big(\varepsilon^\beta[A]_\beta+\varepsilon\|A\|_\infty^2\big),\\
 \|b^r-b\|_\infty
 &\le C_B\big(\varepsilon^\beta[b]_\beta+\varepsilon\|A\|_\infty\|b\|_\infty\big).
 \end{aligned}
 \label{eq:recon}
\end{equation}
Here $[\cdot]_\beta$ is the H\"older seminorm on the four-torus. The constants are uniform in $\varepsilon$ at fixed physical volume.
\end{lemma}

\begin{proof}
Work on the covering space $\mathbb R^4$, with vertices $x_i=\varepsilon i$. The smooth interpolation functions $\rho_i(x/\varepsilon)$ form a partition of unity. At each point only a bounded number of these functions are nonzero. Craig uses the factors $\phi_i=\sqrt{\rho_i}$ to embed a colour vector in the direct sum of the corresponding vertex colour spaces. For each pair of vertices $i,j$ contributing at this point, average the parallel transports from $j$ to $i$ over the 24 orderings of the coordinate directions. Taking the unitary polar factor of this average and removing its determinant phase gives $u_{ij}\in SU(5)$, with $u_{ji}=u_{ij}^\dagger$.

Set $T_{ij}=\phi_i u_{ij}\phi_j$. At $A=0$ this block matrix equals the rank-five orthogonal projector $FF^\dagger$, where $F_i=\phi_i\mathbf1$ and $F^\dagger F=1$. The paths have length $O(\varepsilon)$, so $\|T-FF^\dagger\|\le C\varepsilon B$. Let $\Pi$ be the spectral projector of $T$ onto the cluster of five eigenvalues near one. The resolvent identity on $|\zeta-1|=1/2$ gives $\|\Pi-FF^\dagger\|\le C'\varepsilon B$. The columns of
\begin{align}
 C=\Pi F(F^\dagger\Pi F)^{-1/2}
\end{align}
therefore form a smooth orthonormal basis of the reconstructed fibre. Translation by a torus period relabels the vertices, so this frame descends to the torus.

It remains to fix the determinant of the colour frame. Craig prescribes a unit element $\sigma$ of the fifth exterior power of the fundamental colour fibre. His local frames, obtained by transporting from a vertex and then projecting and orthonormalizing, differ from $F$ by $O(\varepsilon B)$. The normalized weighted average of their exterior products is $\sigma$, so $\wedge^5 C/\sigma=1+O(\varepsilon B)$. Multiplying the columns of $C$ by the principal inverse fifth root of this ratio gives the prescribed volume element. Craig's $SU(5)$ gauge potential in the corrected frame is $(C^\dagger d_x C)_0$, where the subscript denotes the traceless part. Indeed, multiplying $C$ by a scalar unitary matrix adds a multiple of the identity to $C^\dagger d_x C$. Its traceless part is unchanged along the whole family, so differentiating this expression also accounts for the determinant normalization in $b^r$.

Differentiate first at the vacuum along $tA$. For $X_{ij}=\partial_t u_{ij}|_{t=0}$, antisymmetry gives $F^\dagger\dot T F=\sum_{ij}\rho_i\rho_jX_{ij}=0$. Differentiating the spectral projection then yields $\dot\Pi F=\dot T F$ and $\dot C_i=\phi_i\sum_j\rho_jX_{ij}$. Substitution in the connection formula gives
\begin{equation}
 \dot A^r=\sum_{ij}\rho_i\,d_x\rho_j\,X_{ij}.
 \label{eq:linearrec}
\end{equation}
At first order, the line integral of $A$ along each path is traceless and anti-Hermitian. Taking the unitary polar factor and removing its determinant phase both preserve the linear term in the path average. Replacing $A$ on a path by its value at $x$, with the link convention of section~\ref{sec:result}, gives
\begin{align}
 X_{ij}=(x_j-x_i)^\mu A_\mu(x)+O(\varepsilon^{1+\beta}[A]_\beta).
\end{align}
The interpolation functions satisfy $\sum_j\rho_j(x)x_j=x$. In one dimension, the convolution of two centred unit boxes is the linear hat function, which reproduces affine functions under integer translates. Convolution with the remaining centred box and the even mollifier of integral one preserves the first moment. Taking tensor products gives the asserted identity in four dimensions. Its derivative and $\sum_jd_x\rho_j=0$ show that the leading term in \eqref{eq:linearrec} is exactly $A(x)$. Since $|d_x\rho_j|\le C/\varepsilon$, the linear error is $O(\varepsilon^\beta[A]_\beta)$. The same calculation applies to $b$.

Within each cell write the connection as $A^r_\mu=\varepsilon^{-1}f_\mu(x/\varepsilon,\ell)$. Its argument $\ell=(\log U_e)_e$ collects the logarithms of a bounded number of nearby links. The function vanishes at $\ell=0$ and its first two link derivatives are uniformly bounded. The interpolation functions have bounded derivatives in the reference cell, and the spectral gaps keep the resolvents and inverse square roots smooth with uniform bounds. Let $\Lambda(A)$ be the list of integrals of $A$ along these links, in the sign convention of \eqref{eq:linearrec}. Expanding the path-ordered exponentials and taking the logarithms gives $\ell=\Lambda(A)+O(\varepsilon^2\|A\|_\infty^2)$ in the maximum norm over links. Differentiating the series gives $\delta_b\ell=\Lambda(b)+O(\varepsilon^2\|A\|_\infty\|b\|_\infty)$. Taylor's formula bounds the nonlinear part of $f_\mu$ by $C\|\ell\|^2$, and $\|df_\mu(\ell)-df_\mu(0)\|\le C\|\ell\|$ bounds its derivative. Restoring the factor $\varepsilon^{-1}$ from differentiation in $x$ and adding the linear errors proves both estimates in \eqref{eq:recon}.
\end{proof}

\paragraph{Comparison of the two frames.}
The distinction between the identifications can be seen directly from
their defining maps. In the periodic frame of the original bundle put
\begin{align}
 (J_A(x))_i=\phi_i(x)T_A(x_i\leftarrow x),
 \quad
 C_F=\Pi F(F^\dagger\Pi F)^{-1/2},
 \quad
 C_H=\Pi J_A(J_A^\dagger\Pi J_A)^{-1/2},
\end{align}
where $T_A(x_i\leftarrow x)$ is parallel transport along the straight
segment from $x$ to $x_i$. The first identification uses only the
sampled links and the fixed map $F$; the second also uses $A$ between
the vertices. After the determinant corrections described above and those of Hsu
(formula~(F.8) in \cite{Hsu:2026gmi}), both frames have the prescribed colour volume
element. We use the same symbols for these corrected frames. Their relative matrix $g=C_F^\dagger C_H$ is therefore a smooth
periodic $SU(5)$ transformation. If $A_F^r$ and $A_H^r$ denote
the corresponding potentials, then
\begin{equation}
 A_H^r=g^{-1}A_F^r g+g^{-1}d_xg,\qquad
 b_H^r=g^{-1}b_F^r g+d_{A_H^r}\xi_b,\qquad
 \xi_b=g^{-1}\delta_b g .
 \label{eq:recframes}
\end{equation}
Here $b_F^r=\delta_b A_F^r$, $b_H^r=\delta_b A_H^r$, and
$d_{A_H^r}\xi=d_x\xi+[A_H^r,\xi]$. In particular, the tangent is
not obtained by conjugation alone when the change of frame depends on
the background.

Gauge invariance makes the response to the last term in \eqref{eq:recframes} vanish. The estimates below are proved in $C_F$, not as ordinary-derivative bounds in $C_H$; this suffices to compare the scalar weights.

The vacuum polarization is sensitive to derivatives of the gauge field in all four directions. To control it under reconstruction, we first show that the map linearized at $A=0$ reproduces gauge potentials whose components are affine functions of $x$.

\begin{lemma}\label{lem:affine}
Let $R_a=dA^r|_{A=0}$ in the periodic frame of lemma~\ref{lem:reconstruction}.
For smooth $b$ on the unit torus and ordinary spacetime derivatives of order $j=0,1,2$,
\begin{align}
 \|\partial^j(R_a b-b)\|_\infty
 \le C\varepsilon^{2-j}\|b\|_{C^2}.
\end{align}
\end{lemma}
\begin{proof}
The linear map in \eqref{eq:linearrec} is exact on one-forms whose components are affine functions on the covering space. Indeed, the one-dimensional interpolation function $w$ has Fourier transform
$\widehat w(\xi)=\operatorname{sinc}(\xi/2)^3\widehat\psi(\xi)$,
where $\operatorname{sinc}z=\sin z/z$ and $\psi$ is the smooth, even function of unit integral used in the convolution. The transform and its first two derivatives vanish at every nonzero integer multiple of $2\pi$. Periodizing the moments of $w$ and taking tensor products therefore gives, as in the first-moment identity underlying equation~(D.38) in \cite{Hsu:2026gmi},
\begin{align}
 \sum_i\rho_i=1,\qquad \sum_i\rho_i(x_i-x)=0,\qquad
 \sum_i\rho_i(x_i-x)^\mu(x_i-x)^\nu=\varepsilon^2 \Theta^{\mu\nu}
\end{align}
with constant $\Theta$. Write $z_i=x_i-x$. For affine $b$ the linearized path average is
\begin{align}
 X_{ij}=(x_j-x_i)^\mu
 \left(b_\mu(x)+\tfrac12(z_i+z_j)^\nu\partial_\nu b_\mu\right).
\end{align}
Among the 24 orderings of the path segments, each coordinate other than $\mu$ takes its initial value in half the orderings and its final value in the other half while the path traverses the $\mu$-segment. The coordinate along that segment averages to its midpoint value. This proves the displayed identity, including the cases of vanishing displacements. In \eqref{eq:linearrec} the gradient correction becomes
$\frac12\partial_\nu b_\mu\sum_j d\rho_j\,z_j^\mu z_j^\nu=0$
by differentiation of the second moment and the vanishing first moment.

Fix $x$ and subtract the affine Taylor polynomial of $b$ at $x$. Hold this polynomial fixed when differentiating the interpolation operator. The difference is $O(\varepsilon^2\|b\|_{C^2})$ throughout the neighbourhood of the links entering the reconstruction at $x$, and its line integrals along these links are $O(\varepsilon^3\|b\|_{C^2})$. Derivatives of $\rho_i d\rho_j$ of order up to $j$ are bounded by $C\varepsilon^{-1-j}$. This proves all three estimates.
\end{proof}

\paragraph{Fourier components of the reconstructed gauge field.}
Let $\mathcal X$ be the space in which the Fourier coefficients are absolutely summable with weight
$\langle h\rangle(1+\log\langle h\rangle)$, where $\langle h\rangle=1+|h|$.
Even a single plane wave can generate additional Fourier components under reconstruction, because the interpolation functions depend on position within each lattice cell. We estimate these components, allowing arbitrary external momenta. The norms $\mathcal W_1$ and $\mathcal X$ will account for the momentum dependence of the continuum vertices.

\begin{lemma}\label{lem:foldedrecon}
Let $E(A)=A^r(A)-A$ in the frame of lemma~\ref{lem:reconstruction}.
Set $m_{\mathcal W_j}=\varepsilon^{-j}$ for $j=0,1,2$ and $m_{\mathcal X}=\varepsilon^{-1}\ell_\varepsilon$.
Write $A_j=\|A\|_{\mathcal W_j}$ and similarly for $b$. For sufficiently small $\varepsilon A_0$, each of these four output norms satisfies
\begin{align}
 \|E(A)\|_Y
 &\le C m_Y\varepsilon^2 (A_2+A_0 A_1+A_0^3),\nonumber\\
 \|dE(A)[b]\|_Y
 &\le C m_Y\varepsilon^2(b_2+A_0 b_1+A_1 b_0+A_0^2 b_0).
 \label{eq:foldedrecon}
\end{align}
The constant is independent of the Fourier supports. On smooth fields with $\|A\|_{C^2}\le B$, the following estimates hold for $j=0,1,2$ and all sufficiently small $\varepsilon$:
\begin{equation}
 \|E(A)\|_{C^j}\le C_B\varepsilon^{2-j},\qquad
 \|dE(A)[b]\|_{C^j}\le C_B\varepsilon^{2-j}\|b\|_{C^2}.
 \label{eq:mixedrecon}
\end{equation}
\end{lemma}

\begin{proof}
For constant $A_\mu$, the coefficient of $s^2$ in $A^r(sA)$ vanishes even if the components do not commute. To see this, fix $x_0$ and set $B_i=(x_i-x_0)^\mu A_\mu$. These matrices are held fixed when differentiating with respect to $x$. Averaging the path-ordered products over the coordinate directions gives
\begin{align}
 u_{ij}=1+s(B_j-B_i)+\tfrac12s^2(B_j-B_i)^2+O(s^3).
\end{align}
Each pair of distinct directions occurs equally often in either order. The exponential along each straight segment supplies the term quadratic in its own displacement. This expansion is unitary with determinant one through degree two, so both normalizations preserve it. Its quadratic coefficient differs from that of $e^{-sB_i}e^{sB_j}$ by $\tfrac12[B_i,B_j]$.

For the comparison transporters $e^{-sB_i}e^{sB_j}$, which compose exactly, the matrix $T$ is the orthogonal projection onto the range of the isometry with blocks $(\phi_i e^{-sB_i})_i$. The polar frame used above has blocks $\phi_i e^{-sB_i} H$, where $H$ is the unitary polar factor of $\sum_j\rho_j e^{sB_j}$. Its connection is $(H^\dagger dH)_0$. The quadratic coefficient of $\sum_j\rho_j e^{sB_j}$ is Hermitian, so expanding the unitary polar factor gives $\log H=s\sum_j\rho_j B_j+O(s^3)$. At $x_0$ the first moment gives $\sum_j\rho_j B_j=0$, and hence the quadratic coefficient of $H^\dagger dH$ vanishes there. The difference between the two sets of transports starts at order $s^2$, so its contribution at this order is given by the linearization \eqref{eq:linearrec}. It is $\tfrac12[\sum_i\rho_i B_i,\sum_jd\rho_j B_j]=0$ at the same point. Since $x_0$ was arbitrary, this proves the cancellation in the chosen frame.

Let $R_m[u_1,\ldots,u_m]$ be the $m$-th derivative of the reconstructed field at the vacuum divided by $m!$. We first bound it locally for arbitrary smooth inputs. For a cell $Q$, let $Q^+$ contain all edges used in the reconstruction on $Q$. This neighbourhood has diameter $O(\varepsilon)$. Write $y=x/\varepsilon$ for the rescaled coordinates. For each fixed $J$,
\begin{equation}
 \max_{|\nu|\le J}\sup_{\varepsilon y\in Q}
 \big|\partial_y^\nu \{R_m[u_1,\ldots,u_m](\varepsilon y)\}\big|
 \le C_J^m\varepsilon^{m-1}\prod_{i=1}^m\|u_i\|_{L^\infty(Q^+)} .
 \label{eq:cellcoefficient}
\end{equation}
To see this, insert $\sum_i t_i u_i$ into the path-ordered exponentials. All edge matrices remain close to the identity for $\varepsilon\sum_i |t_i|\|u_i\|_{L^\infty(Q^+)}<r$, even for complex parameters. For real inputs the adjoint of each path product is the product along the reversed path. This gives its holomorphic extension in the parameters, including in the polar factors. The spectral contour stays fixed and the inverse square roots and determinant roots are analytic on this neighbourhood. The edge matrices are constant in $y$ within a cell, so cell derivatives act on the smooth copying functions and the matrix functions built from them. For fixed $J$ the resulting cell formula is uniformly bounded by $C_J/\varepsilon$. The power $\varepsilon^{-1}$ comes from $d_x C$ in the connection. Cauchy's estimate on circles of radii $r/(2m\varepsilon\|u_i\|_{L^\infty(Q^+)})$ proves \eqref{eq:cellcoefficient} since $m^m/m!\le e^m$. An input that vanishes on $Q^+$ gives zero by locality.

We can now prove \eqref{eq:mixedrecon} using just two bounded derivatives of the inputs. Lemma~\ref{lem:affine} deals with the linear term. For the quadratic term, freeze each input at the centre of $Q$, keeping this value fixed when differentiating in $y$. Polarization of the constant-field cancellation gives $R_2[u_{\rm c},v_{\rm c}]=0$ everywhere for any pair of constant potentials. On $Q^+$, $\|u-u_{\rm c}\|_\infty\le C\varepsilon\|\nabla u\|_\infty$. By bilinearity and \eqref{eq:cellcoefficient}, the $C^j$ norm in physical coordinates of $R_2[u,v]$ on $Q$ is at most
\begin{align}
 C\varepsilon^{2-j}\big(\|\nabla u\|_\infty\|v\|_\infty
                    +\|u\|_\infty\|\nabla v\|_\infty\big).
\end{align}
For $m\ge3$, restoring physical derivatives costs at most $\varepsilon^{-j}$. The sum of \eqref{eq:cellcoefficient} at these orders is $O_B(\varepsilon^{2-j})$. A derivative in the direction $b$ replaces one input by $b$ and introduces a factor $m$, which leaves the geometric series summable. Together these bounds prove \eqref{eq:mixedrecon} in the global frame, including on cell faces.

For the Fourier estimates, take plane waves of unit amplitude with momenta $h_i$, extending the real multilinear coefficients to complex inputs. Translation by $\varepsilon n$, $n\in\mathbb Z^4$, multiplies the output by $\exp(i\varepsilon n\cdot\sum h_i)$. Let $h_*$ be the representative of this total momentum in the first Brillouin zone. Then
\begin{align}
 R_m[u_1,\ldots,u_m](x)=e^{ih_*\cdot x}g_m(x/\varepsilon),
 \qquad g_m(y+n)=g_m(y).
\end{align}
Across a cell face a translation merely relabels the vertices in the smooth reconstruction. Removing the plane wave preserves the bound $\|g_m\|_{C^J}\le C_J^m\varepsilon^{m-1}$ from \eqref{eq:cellcoefficient}, since $\varepsilon h_*$ is bounded. The output momenta are $h_*+2\pi n/\varepsilon$. Their weights in $\mathcal W_j$ are at most $C\varepsilon^{-j}(1+|n|)^j$ for $0\le j\le2$, and their weights in $\mathcal X$ are at most $C\varepsilon^{-1}\ell_\varepsilon(1+|n|)(1+\log(2+|n|))$. For instance, $J=8$ gives enough Fourier decay in four dimensions to sum all four weights. Thus
\begin{align}
 \|e^{ih_*\cdot x}g(x/\varepsilon)\|_Y\le C m_Y\|g\|_{C^8}.
\end{align}

For one input with $\varepsilon |h|<c$, subtract its affine Taylor polynomial about a fixed point of the cell. The first reconstruction coefficient reproduces that polynomial exactly by lemma~\ref{lem:affine}. On all links entering the cell the remainder is bounded by $C(\varepsilon|h|)^2$. This also bounds the periodic factor of the error in $C^J$. For the reconstructed term, the derivatives act only on the interpolation functions. For the original plane wave, the rescaled derivatives of the Taylor remainder are $O_J((\varepsilon|h|)^2)$. Multiplication by $e^{-i\varepsilon h_*\cdot y}$ preserves this estimate. For $\varepsilon|h|\ge c$ the reconstructed wave has norm at most $Cm_Y$, and the norm of the original wave equals its Fourier weight in $Y$. Both are bounded by $Cm_Y\varepsilon^2\langle h\rangle^2$. For $Y=\mathcal X$ this follows because $(1+\log t)/t$ decreases on $[1,\infty)$.

For two inputs with $\varepsilon (|h_1|+|h_2|)<c$, write each field as its value at the cell centre plus a remainder. The coefficient with two constant fields vanishes by polarization of the constant-field identity above. Every term that remains contains at least one input of size $O(\varepsilon(|h_1|+|h_2|))$ on the links entering the cell. Thus the periodic part of the quadratic coefficient is bounded by $C\varepsilon^2(\langle h_1\rangle+\langle h_2\rangle)$ in $C^J$. For larger momenta this follows directly from the uniform quadratic bound $C\varepsilon$. Finally sum the bounds for $m\ge3$ against the absolutely summable input coefficients, using $\sum_{m\ge3}C^m\varepsilon^{m-1}A_0^m\le C'\varepsilon^2 A_0^3$. Differentiating the series replaces one of the $m$ fields by $b$. The extra factor of $m$ preserves convergence and gives the second estimate in \eqref{eq:foldedrecon}.
\end{proof}

\paragraph{A mode invisible to every lattice link.}
The three powers in \eqref{eq:mixedrecon} cannot be improved uniformly
over the stated $C^2$-bounded class. On the unit torus set
$\varepsilon=N^{-1}$, $h_N=2\pi N$, and choose a fixed nonzero
anti-Hermitian generator $T$. Consider
\begin{equation}
 b_N=h_N^{-2}T\sin(h_Nx^1)\,dx^2,\qquad A_N(t)=t b_N .
 \label{eq:invisiblemode}
\end{equation}
Every edge in direction two lies at $x^1=n/N$, where this potential
vanishes. Its components along all other edges are zero. Hence
$U_a(A_N(t))=1$ for every $t$. The reconstructed connection in
our frame is $A_F^r=0$, and its $t$ derivative is zero as well.
Nevertheless, the original curvature is
$t h_N^{-1}T\cos(h_Nx^1)\,dx^1\wedge dx^2$, so this is not a
pure gauge field for $t\ne0$. Writing $E=A_F^r-A$, we obtain
\begin{equation}
 \|\partial_1^j E(A_N(t))\|_\infty
      =|t|\,\|T\|h_N^{j-2},
      \quad
 \|\partial_1^j dE(A_N(t))[b_N]\|_\infty
      =\|T\|h_N^{j-2},\quad j=0,1,2.
 \label{eq:invisibleorders}
\end{equation}
For $t$ in a fixed bounded interval the backgrounds and tangents have
uniformly bounded $C^2$ norms. Since $\|A_N(t)\|_\infty=O(N^{-2})$,
the free Dirac gap also persists uniformly for sufficiently large
$N$. For any fixed $t\ne0$, the errors attain the orders $\varepsilon^2$, $\varepsilon$, and
$1$, including for the tangent. This establishes the sharpness of the
uniform reconstruction bounds in this frame; it neither concerns a
single fixed smooth background nor supplies a lower bound on the
fermion phase error.

For $n=2,3,4$, the continuum vacuum vertices $d^n\mathcal M(0)$ in momentum space are bounded in norm by $C H^{4-n}(1+\log H)$ (equation~(D.28) in \cite{Hsu:2026gmi}), where $H=1+\sum_i|h_i|$. In the traces regularized with the free operator in \eqref{eq:BF}-\eqref{eq:weight}, the $n$ propagators carry shifted loop momenta $k+p_i$, with $|p_i|\le H$. On $|k|\le 2H$, H\"older's inequality gives the asserted power from $\sum_{|k|\le 3H}|k|^{-n}$, with a logarithm for $n=4$. The external bosonic momenta preserve the fermion momentum grid, and the spin structure excludes zero from this grid. Above $2H$, Taylor subtraction in the momenta through degree $4-n$ leaves a summable kernel bounded by $C H^{5-n}|k|^{-5}$. For each Taylor term of degree $d\le4-n$, subtract the sum over $|k|\le2H$ from its finite part on the full fermion momentum grid. The difference is bounded by $C H^d(1+\sum_{|k|\le2H}|k|^{-n-d})$. The local residue has at most $4-n$ derivatives and obeys the same bound.

For $n=3,4$ the resulting momentum bound $CH^{4-n}(1+\log H)$ gives
\begin{align}
 |d^n\mathcal M(0)[u_1,\ldots,u_n]|\le C\sum_i\|u_i\|_{\mathcal X}\prod_{j\ne i}\|u_j\|_{\mathcal W_0}.
\end{align}
For $n=2$ momentum conservation pairs $h$ with $-h$, giving
$|d^2\mathcal M(0)[u,v]|\le C\|u\|_{\mathcal X}\|v\|_{\mathcal W_1}$, also with the two norms exchanged. For $\|A\|_{\mathcal W_3}\le B$, insert $A^r=A+E$ and $b^r=b+dE(A)[b]$ into the continuum polynomial $L_c$ through fourth order in the fields. In a term containing one reconstruction error, momentum conservation bounds the magnitude of its momentum by the sum of the magnitudes of the other momenta. It follows that the entire vertex factor $CH^2(1+\log H)$ can be bounded using the $\mathcal W_3$ norms of the other fields. The single error is measured in $\mathcal W_0$, where it is $O_B(\varepsilon^2)$, with an additional factor of $\|b\|_{\mathcal W_3}$ when the variation occurs in the error. If both fields in the two-point vertex are errors, use $\mathcal X$ for one and $\mathcal W_1$ for the other. Their product is $O_B(\varepsilon^2\ell_\varepsilon)$. For a three- or four-point vertex with at least two errors, each term of the sum over $i$ above contains at least one error measured in $\mathcal W_0$. The field measured in $\mathcal X$ has bounded norm if it is an original field and norm $O_B(\varepsilon\ell_\varepsilon)$ if it is a reconstruction error. Keeping the variation norm throughout gives

\begin{align}
 |L_c(A^r;b^r)-L_c(A;b)|\le C_B\varepsilon^2\ell_\varepsilon\,\|b\|_{\mathcal W_3}.
\end{align}
The same argument applied to individual Taylor coefficients gives this bound for the reconstructed vacuum vertices through degree four. The one-point vertex vanishes by the colour trace.

Finally, suppose the original fields have Fourier support in $|h|\le M$, with $M\ge1$. Assume their $\mathcal W_0$ norms are bounded and $\varepsilon M$ is small. By \eqref{eq:foldedrecon} a reconstruction error has norm at most $C_B\varepsilon^2 M^2$ in $\mathcal W_0$, $C_B\varepsilon M^2$ in $\mathcal W_1$, and $C_B\varepsilon\ell_\varepsilon M^2$ in $\mathcal X$, with a factor $\|b\|_{\mathcal W_0}$ if the field is varied. With only one such error, the other fields impose $H\le CM$ by momentum conservation, so the vertex is bounded by $CM^2(1+\log M)$. With two errors in the two-point vertex, the $\mathcal X$ and $\mathcal W_1$ bounds give $C_B\varepsilon^2\ell_\varepsilon M^4$.

For the three- and four-point vertices with at least two errors, there are two possibilities in the sum over $i$. If $u_i$ is an error, its $\mathcal X$ norm times the $\mathcal W_0$ norm of another error is bounded by $C_B\varepsilon^3\ell_\varepsilon M^4$. If $u_i$ is an original field, its $\mathcal X$ norm times the unweighted norms of two errors is bounded by $C_B M(1+\log M)\varepsilon^4 M^4$. All remaining unweighted factors are bounded. Since $\varepsilon M$ is small, these estimates give
\begin{align}
 |L_c(A^r;b^r)-L_c(A;b)|\le C_B\varepsilon^2\ell_\varepsilon M^4\|b\|_{\mathcal W_0}.
\end{align}

\section{Vacuum polarization in infinite and finite volume}
\label{app:vacuum}

After subtraction of the local gauge kinetic term, two symmetries give a quadratic convergence rate, up to a logarithmic factor, for the lattice vacuum polarization. Reversing the direction of the lattice step leaves $D_a^\dagger D_a$ invariant and eliminates the first-order error in its vertices, and inverting all four coordinates cancels the cubic term in the external-momentum expansion after integration over the loop momentum. We compare the two polarizations first in infinite volume and then on the four-torus, with a bound that remains uniform when the external momentum grows as $a$ decreases.

\paragraph{The one-loop integrand.}
Let $k$ be the fermion loop momentum and $h,-h$ the momenta of two external gauge fields. At $A=0$, take the second variation of the logarithm of the modulus of the lattice Weyl determinant. For fixed colour and vector components of the external fields, let $I_\varepsilon(k,h)$ denote the loop integrand per unit volume. The chiral multiplicity identity in section~\ref{sec:two} expresses this Hessian as the second variation of $\tfrac14\Tr\log Q_\varepsilon$, with $Q_\varepsilon=D_a^\dagger D_a$. There is a contact diagram with a two-field vertex and a free propagator, giving $Q_{\varepsilon,0}^{-1}\delta^2Q_\varepsilon$, and a bubble diagram with two one-field vertices separated by free propagators $Q_{\varepsilon,0}^{-1}$. Here the second subscript $0$ means $A=0$. We assign the same loop momentum to the same propagator in the lattice and continuum diagrams.

For $\varepsilon(1+|h|)<c$ and $|k|>2(1+|h|)$, each diagram satisfies
\begin{equation}
 |\partial_k^m\partial_h^n I_\varepsilon(k,h)|\le C_{m,n}|k|^{-2-m-n}.
 \label{eq:bubblesymbol}
\end{equation}
The constants are uniform in $h$ at the derivative orders used below. The contour representation in appendix~\ref{app:vertices} gives smooth periodic functions of the dimensionless fermion momenta for the free overlap operator and its first two link derivatives. In units of the torus size, $Q_{\varepsilon,0}^{-1}=O(|k|^{-2})$, $\delta Q_\varepsilon=O(|k|)$ and $\delta^2Q_\varepsilon=O(1)$ in this momentum range. Differentiating with respect to $k$ introduces a factor $\varepsilon\le C/|k|$ in any smooth function of the dimensionless momentum $\varepsilon k$. Differentiating the free propagator improves its decay by one inverse power of $|k|$. These bounds also hold across identified faces of the Brillouin zone. The continuum bounds follow from the free symbol $Q_{0,0}(k)=D_0(k)^\dagger D_0(k)=|k|^2\mathbf 1$.

\paragraph{Matching the polarization tensor.}
Write $D_{2,a}(h,-h)$ for the second variation at zero field of $\mathcal M_a-\mathcal M-c_R I$, evaluated on these two modes of the original gauge potential, as in section~\ref{sec:ward}.
\begin{lemma}\label{lem:evenbubble}
Let $h\in2\pi\mathbb Z^4$ and $H=1+|h|$. On the torus with the chosen spin structure there are constants $c,C>0$ such that for $\varepsilon H<c$ and unit external polarizations,
\begin{equation}
 |D_{2,a}(h,-h)|\le C\varepsilon^2\ell_\varepsilon H^4 .
 \label{eq:anchor-rate}
\end{equation}
\end{lemma}

\begin{proof}
Consider parallel transport along a signed displacement $t\hat\mu$ on the covering space. The physical lattice spacing in units of $L$ is $\varepsilon>0$. On the fixed space of spinors on the cover, the covariant translation $T_{t,\mu}\psi(x)=U(x,x+t\hat\mu)\psi(x+t\hat\mu)$ satisfies $T_{-t,\mu}=T_{t,\mu}^\dagger$ for a unitary connection. The dimensionless Wilson kernel is
\begin{align}
 X_t=-1+\sum_\mu\big[1-(T_{t,\mu}+T_{t,\mu}^{\dagger})/2
                  +\gamma_\mu(T_{t,\mu}-T_{t,\mu}^{\dagger})/2\big].
\end{align}
Thus $X_{-t}=X_t^\dagger$. The gap of the Wilson kernel near the vacuum ensures that its unitary polar factor $U_t=X_t(X_t^\dagger X_t)^{-1/2}$ exists and obeys $U_{-t}=U_t^\dagger$. Since $D_t=(1+U_t)/t$ is normal, we have
\begin{equation}
 D_{-t}=-D_t^\dagger,\qquad Q_{-t}=Q_t,\qquad Q_t=D_t^\dagger D_t .
 \label{eq:signedstep}
\end{equation}
Differentiation with respect to the external field and multilinear extension from real gauge fields show that every Fourier matrix element of $\delta Q_t$ and $\delta^2 Q_t$ at the vacuum is even in $t$. This includes the contact vertex. The argument on the cover allows continuous external momenta. Smoothness in $t$ is needed only for these matrix elements and the free Fourier symbol. It follows from the fixed contour in appendix~\ref{app:vertices}. In the following estimates $Q_0(A)=D_A^\dagger D_A$ denotes the continuum operator.

Write $r=|k|$ and $s=|k+h|$ for one assignment of the loop momentum. At $\varepsilon r,\varepsilon s<c_0$, expansion about the zero of the free Dirac symbol, together with evenness in the step, gives
\begin{align}
 Q_{\varepsilon,0}(k)=r^2+O(\varepsilon^2 r^4),\qquad
 Q_{\varepsilon,0}(k)^{-1}-r^{-2}=O(\varepsilon^2).
\end{align}
To bound the one-field vertex, measure all momenta in units of $\bar r=r+s>0$. The free symbol $D_t(k)/\bar r$ is smooth in $t\bar r$ and $k/\bar r$ on this bounded set. The apparent pole at $t\bar r=0$ is removable because the free dimensionless symbol vanishes at zero momentum. In $\delta D_t$ the edge integral cancels the factor $1/t$, and the contour resolvents stay bounded. At $t=0$, the limits of $D_t(k)/\bar r$ and $\delta D_t$ are $i\gamma\cdot k/\bar r$ and the linear continuum vertex, respectively. The resulting matrix elements of $\bar r^{-1}\delta Q_t$ have bounded derivatives in $t\bar r$, even when either internal momentum vanishes. Since they are even in $t\bar r$,
\begin{align}
 |\delta Q_\varepsilon-\delta Q_0|\le C\varepsilon^2(r+s)^3,\qquad
 |\delta Q_\varepsilon|+|\delta Q_0|\le C(r+s).
\end{align}
For the two-field vertex in the ball $r<4H$, rescale by $H$. Differentiating $D_t^\dagger D_t$ twice gives either a product of two one-field vertices or a product of the free symbol and a two-field vertex of $D_t$. Each product is smooth in $tH$, since the latter vertex carries an explicit factor of $t$. Evenness gives $|\delta^2 Q_\varepsilon-\delta^2 Q_0|\le C\varepsilon^2 H^2$.

With $I_0$ denoting the continuum integrand, these estimates give
\begin{align}
 |I_\varepsilon-I_0|\le C\varepsilon^2H^2(r^{-2}+s^{-2})
\end{align}
in the inner ball for both the contact term and the bubble with this momentum assignment. For example, replacing one vertex in the bubble gives at most $C\varepsilon^2(r+s)^4/(r^2s^2)$. Use $r+s\le9H$ and $(r+s)^2/(r^2s^2)\le2(r^{-2}+s^{-2})$. Replacing a free propagator gives $O(\varepsilon^2)$ times the other propagator and the two vertices. Interchanging the external legs replaces $s$ by $|k-h|$. For every real external momentum in infinite volume, integrating either inverse-square factor over the ball gives $O(H^2)$. On the torus an allowed bosonic external momentum translates the set of fermion momenta into itself, so its discrete sum has the same bound. The inner region therefore contributes $O(\varepsilon^2H^4)$, including at $h=0$, where the poles coincide.

In the range $2H<r<c_0/\varepsilon$, the internal momenta divided by $r$ stay in a compact annulus as the external momentum varies from zero to $h$. Holding $r$ fixed in this rescaling, $r^2 I_t$ and its derivatives with respect to the rescaled momenta are smooth even functions of $tr$. Taylor expansion at $t=0$ gives
\begin{equation}
 |\partial_k^m\partial_h^j(I_\varepsilon-I_0)|\le C_{m,j}\varepsilon^2r^{-m-j}.
 \label{eq:improved-bubble}
\end{equation}
At $h=0$ this estimate for the derivatives holds for every $0<r<c_0/\varepsilon$. For $r>c_0/\varepsilon$, choose $c\ll c_0$ and apply \eqref{eq:bubblesymbol} to the two regularizations separately. Shifts of length at most $|h|$ keep the internal momenta a distance of order $r$ from the nearest pole, including at the faces of the Brillouin zone.

We first perform the subtraction in infinite volume. Subtract the Taylor polynomial in the external momentum through cubic order, multiplying it by a fixed smooth function of $r$ that vanishes near zero and equals one for $r>2$. Inversion of all four coordinates reverses both external one-forms, so the cubic Taylor coefficient is odd in $k$. Its integral is zero in either regularization. In the intermediate annulus, the fourth external derivative in \eqref{eq:improved-bubble} bounds the integral of the difference of the remainders by
\begin{align}
 C\varepsilon^2H^4\int_{2H}^{c_0/\varepsilon}r^{-1}\,dr.
\end{align}
In the inner ball the original integrands obey the estimate above. The difference of the Taylor terms there contributes at most $C\varepsilon^2\sum_{j=0}^3 H^j\int_0^{4H}r^{3-j}\,dr=O(\varepsilon^2H^4)$. In the ultraviolet region beyond $c_0/\varepsilon$, the integrals of the individual remainder bounds are at most $C H^4\int_{c_0/\varepsilon}^\infty r^{-3}\,dr=O(\varepsilon^2H^4)$. The local heat-kernel subtractions in the continuum contribute only to the polynomial, as justified below. We have thus expressed the infinite-volume polarization difference as a polynomial of degree at most two plus a remainder bounded by $C\varepsilon^2H^4[1+\log(1/(\varepsilon H))]$. The cutoff on the Taylor terms is independent of $h$, so the same polynomial applies throughout the stated range.

In infinite volume we can average the polynomial over the hypercubic group. The resulting tensor has the form
\begin{align}
 m\delta_{\mu\nu}+u|h|^2\delta_{\mu\nu}+v h_\mu h_\nu+
 w\delta_{\mu\nu}h_\mu^2 ,
\end{align}
with no sum in the last term. Both exact polarization tensors obey $h_\mu\Pi_{\mu\nu}=0$ as functions of the original continuum gauge potential, by gauge invariance and the vanishing of the one-point function. Contract the polynomial with $h_\mu$ first at two fixed momenta of different lengths along the same axis, then at a momentum with two nonzero components of unequal magnitude. The remainder bound gives $m,w,u+v=O(\varepsilon^2\ell_\varepsilon)$. This leaves a single coefficient multiplying the transverse tensor.

We determine it at momenta along one axis. Work first with one Dirac fermion in the fundamental representation. Fix an axial momentum $h_*\ne0$ and take a unit polarization perpendicular to it. At spacing $t>0$ the transverse lattice entry at $\lambda h_*$, divided by $|\lambda h_*|^2$, is $F(\lambda t)$ in the notation of \eqref{eq:axialdef}. Indeed, in infinite volume the change of loop variable $k\mapsto tk$ leaves a dimensionless integral depending only on $t\lambda h_*$ after this division. Compare the normalized lattice-minus-continuum entries at $h_*$ and $\lambda h_*$ for $1\le\lambda\le2$. The unknown transverse coefficient is the same at both momenta. The residual mass term and the Taylor remainders contribute $O(t^2\ell_t)$ uniformly on this interval, since $|\lambda h_*|$ is bounded above and away from zero. By \eqref{eq:flatquillen} the difference between the two normalized continuum entries, with the larger momentum taken first, is $\log(\lambda^2)/(24\pi^2)$. Hence
\begin{equation}
 \sup_{1\le\lambda\le2}|g(\lambda t)-g(t)|\le Ct^2\ell_t,
 \quad
 g(t)=F(t)-\frac{\log(t^2|h_*|^2)}{24\pi^2},
 \quad
 \ell_t=1+\log(1/t).
 \label{eq:axialosc}
\end{equation}
For any $0<s\le t$ choose $j\ge0$ with $s\le 2^{-j}t<2s$. Applying \eqref{eq:axialosc} at the successive dyadic scales and once between $s$ and $2^{-j}t$ gives
\begin{align}
 |g(s)-g(t)|\le C\sum_{i\ge0}4^{-i}t^2(\ell_t+i\log2)
                  \le C't^2\ell_t .
\end{align}
It follows that the limit defining $k_f(1)$ in \eqref{eq:axialdef} exists and that $g(t)=k_f(1)+O(t^2\ell_t)$. Rescaling the momenta as above and permuting the axes show that the limit is independent of the chosen axial momentum.

To identify this constant with the one tabulated in \cite{Alexandrou:1999wr}, observe that at axial momentum each transverse diagonal entry is one third of the trace over vector indices. The Ward identity sets the longitudinal entry to zero, and hypercubic symmetry makes the three transverse entries equal. For a transverse plane wave on this axis the potential is constant along each link pointing in the polarization direction, so its exact edge transport agrees with the midpoint prescription. The normalization by squared lattice momentum (equation~(2.8) in \cite{Alexandrou:1999wr}) differs from ours by a relative $O(t^2)$. The resulting $O(t^2\ell_t)$ change leaves the intercept unchanged. With the sign of $\log|W|$, equations~(2.8) and~(3.4) in \cite{Alexandrou:1999wr} thus identify this intercept with $+k_f(1)$. Restoring the colour factor $T(R)$ and subtracting \eqref{eq:flatquillen} gives $c_R(a,\mu)+O(\varepsilon^2\ell_\varepsilon)$ for the transverse coefficient. After subtraction of $c_R d^2 I$, the remaining polynomial is bounded by $C\varepsilon^2\ell_\varepsilon H^2$.

The continuum Taylor subtraction can be justified before removing the proper-time regulator. Restrict the heat integral to proper times greater than $\tau>0$. The first derivative contains $f_\tau(Q_{0,0})$ with $f_\tau(x)=e^{-\tau x}/x$. In the Hessian the contact vertex is multiplied by $f_\tau(r^2)$, and the bubble by the divided difference
\begin{align}
 \frac{f_\tau(r^2)-f_\tau(s^2)}{r^2-s^2}
   =\int_0^1 f_\tau'((1-v)r^2+v s^2)\,dv .
\end{align}
Here $s=|k+h|$ or $s=|k-h|$, as in the lattice bubble. Since $|f_\tau'(x)|\le Cx^{-2}$, the divided difference is bounded by $C/(r^2s^2)$, with the bound extending to equal arguments by continuity. For $r>2H$ the required derivatives of $e^{-\tau r^2 x}/x$ on the rescaled annulus are bounded uniformly in $\tau r^2$. At $\tau=0$ the divided difference equals $-1/(r^2s^2)$. The continuum bounds imply dominated convergence in the inner ball and for the Taylor remainder in the outer region. The local heat-kernel counterterms in the Hessian have degree at most two in $h$ and agree per unit volume on the torus and on $\mathbb R^4$. After removing these counterterms and the convergent remainder, evaluate the polynomial at a fixed set of momenta sufficient to determine a quadratic. Convergence at those momenta implies convergence of its coefficients. We can therefore take $\tau\downarrow0$ before applying \eqref{eq:improved-bubble}.

It remains to compare the finite-volume corrections. For each regularization this correction is the torus momentum sum minus the infinite-volume integral. Split the loop momenta at $2H$ and $c_0/\varepsilon$ with smooth cutoffs on the respective scales. In the inner region both the sum and the integral of $I_\varepsilon-I_0$ are $O(\varepsilon^2H^4)$. In the intermediate region the $L^1$ norm of its sixth loop-momentum derivatives is at most $C\varepsilon^2H^{-2}$, including derivatives of the cutoffs. Beyond $c_0/(2\varepsilon)$ the sixth derivatives of each kernel have $L^1$ norms $O(\varepsilon^4)$ by \eqref{eq:bubblesymbol}. Poisson summation expresses each finite-volume correction as a sum over nonzero winding vectors $m\in\mathbb Z^4$. The spin structure gives the same phase in two regularizations. Six integrations by parts give the summable factor $C|m|^{-6}$. Periodicity of the lattice integrand cancels boundary terms on opposite faces of the Brillouin zone. Combining the winding contributions with the infinite-volume estimate proves \eqref{eq:anchor-rate} for every allowed $h$, including zero.
\end{proof}

The factor $\ell_\varepsilon$ in \eqref{eq:anchor-rate} cannot be removed: the one-loop computation of section~\ref{sec:numerics}, confirmed there by the exact linear response, gives the term $-t^2|h_*|^2\log(t^2|h_*|^2)/(720\pi^2)$ in the transverse function \eqref{eq:axialsix}, from the dimension-six part of the chirally symmetric operator \eqref{eq:dchi}.

The case $h=0$ illustrates why finite-volume corrections matter. A constant potential along a single colour generator shifts the loop momentum in each colour weight space. This shift leaves the infinite-volume integral invariant both on $\mathbb R^4$ and on the periodic Brillouin zone. Its second variation consequently vanishes. The colour tensor is proportional to $\delta^{bc}$, so the conclusion extends to two arbitrary constant insertions. On the torus a constant gauge potential can have a nontrivial holonomy. For example, let only the Euclidean time direction be antiperiodic and let $\theta$ be the temporal holonomy angle for a single Weyl fermion of unit charge. For each spatial momentum $2\pi\mathbf n$, the cosine product formula evaluates the product over temporal frequencies and gives the modulus ratio $(\cosh E+\cos\theta)/(\cosh E+1)$, with $E=2\pi|\mathbf n|$. The continuum second derivative is therefore
\begin{align}
 \left.\partial_\theta^2\log |W_{\rm cont}|\right|_{\theta=0}
 =-\sum_{\mathbf n\in\mathbb Z^3}\frac{1}{1+\cosh(2\pi|\mathbf n|)} .
\end{align}
The lattice second derivative converges to this value by lemma~\ref{lem:evenbubble}. The value is strictly negative even though the quadratic gauge action $I_2$ vanishes for this flat connection.

\section{Local terms in three- and four-point functions}
\label{app:localjets}

The lattice and continuum three- and four-point functions both contain contributions that are nonlocal in position space. We show that these contributions agree in the continuum limit. Applied to their difference, Taylor subtraction at zero external momentum isolates a local polynomial and leaves a remainder that tends to zero with the lattice spacing. Antiperiodicity in at least one direction allows the subtraction directly on the torus, even when a partial sum of external momenta vanishes.

\begin{lemma}\label{lem:jetbound}
Let $n=3,4$, $d=4-n$ and $H=1+\sum_j|h_j|$. For each $\varepsilon$ there is a tensor polynomial $P_{n,\varepsilon}$ of degree at most $d$ in the independent external momenta such that the difference of the lattice and continuum $n$-point vertices, evaluated before reconstruction and after subtraction of $P_{n,\varepsilon}$, is bounded by
\begin{equation}
 C\varepsilon H^{5-n}\big(1+\mathbf{1}_{n=4}\log H\big)
 \label{eq:jetbound}
\end{equation}
for unit external polarizations whenever $\sum_j h_j=0$ and $\varepsilon H<c$. The constants depend only on the fixed torus, spin structure, representation and spectral scale, with $c>0$ sufficiently small.
\end{lemma}

\begin{proof}
Assign an independent external field to each leg and set $h_n=-\sum_{i<n}h_i$. The allowed fermion momenta form $\Lambda_\theta=2\pi(\mathbb Z^4+\theta)$, where each component of $\theta$ is either $0$ or $1/2$ and at least one equals $1/2$. Every partial sum of the bosonic external momenta translates this set into itself. Thus each free propagator in the loop is evaluated at a nonzero momentum, even when consecutive propagators carry the same momentum.

Near the vacuum, theorem~\ref{thm:two} identifies the logarithm of the Weyl determinant modulus with half the real part of the full Dirac log determinant, up to an additive constant. Expand $\frac12\operatorname{Re}\Tr\log(1+S_aV_a)$ on real gauge fields and take its $n$-linear coefficient. Complex external plane waves are inserted by multilinear extension. In particular, if $Z(k;h,v)$ is a Dirac loop integrand including the Weyl normalization, with external momenta $h$ and polarizations $v$, its contribution to the modulus vertex is
\begin{equation}
 \tfrac12\big[Z(k;h,v)+\overline{Z(k;-h,\bar v)}\big].
 \label{eq:reallegs}
\end{equation}
Complex conjugation acts here on the components in a real basis for the Lie algebra and for one-forms. It sends $v_i e^{ih_i x}$ to $\bar v_i e^{-ih_i x}$. The same prescription applies in the continuum, first for real values of the spectral parameter in the domain of convergence and then by analytic continuation.

An ordered term in the lattice expansion attaches the $n$ external fields to $m$ vertices around the fermion loop. A vertex with $q$ insertions carries $\varepsilon^{q-1}v_q$, with $v_q$ smooth in the dimensionless momenta. The vacuum contour resolvents in appendix~\ref{app:vertices} are smooth periodic matrices in the loop momentum. The ordered edge integrals supply one factor of the edge length per insertion and an integral of the resulting exponential over a unit simplex. This also bounds the momentum derivatives of $v_q$ to every fixed order for bounded dimensionless external momenta. The free propagator between successive vertices carries the loop momentum shifted by the external momenta already inserted. With the loop momentum $k$ fixed at the first vertex, denote the sum over all ordered attachments by $I_{n,\varepsilon}(k,h)$, including the factors from the logarithm and the permutations of the external fields.

In the continuum, let $Q_D=D_A^\dagger D_A$. The Weyl log modulus has variation
\begin{align}
 \delta\!\left[\tfrac14\log\det_\zeta(Q_D/\bar\mu^2)\right]
 =\tfrac12\operatorname{Re}\tr^{Q_D/\bar\mu^2}(D_A^{-1}\delta D_A).
\end{align}
Replacing the spectral weight by its free value adds the local residue in \eqref{eq:weight}. At order $n$ in the gauge field, counting the variation as one insertion, this residue contains at most $4-n$ spacetime derivatives. Indeed, each field lowers the order of the large-momentum symbol of $D_A^{-1}\delta D_A$ by at least one. In the expansion of $\log Q_D$ in resolvents, the terms in $Q_D-Q_{D,0}$ linear and quadratic in $A$ have differential orders at most one and zero, respectively, and give the same gain per insertion. Differentiating a coefficient in a symbol product entails a paired momentum derivative. A derivative already acting on $A$ in $Q_D-Q_{D,0}$ also lowers the available order. Only the term of order $-4$ contributes to the residue. Integrating along the path $sA$ preserves this bound on the number of derivatives.

The remaining coefficient is the finite part at $z=0$ of
\begin{align}
 \sum_{k\in\Lambda_\theta} (|k|/\bar\mu)^{-2z} I_{n,0}(k,h).
\end{align}
Here $I_{n,0}$ is the loop with $n$ linear gauge vertices. Taking the varied field at the last vertex and integrating $s^{n-1}$ from zero to one gives the factor $1/n$ in the logarithm. With the same loop-momentum assignment in the two regularizations, $I_{n,\varepsilon}\to I_{n,0}$ at every fixed allowed $k$. The contact vertices with two or more insertions vanish in this limit.

For each fixed loop momentum $k$, expand the two integrands in the independent external momenta about zero through degree $d$. The terms added back after subtraction, together with the local residue, give $P_{n,\varepsilon}$. The continuum Taylor coefficients are homogeneous symbols in $k$, whose spectrally weighted sums on $\Lambda_\theta$ have meromorphic continuation. The Taylor remainder is absolutely summable at $z=0$. Take $c_0$ small, with $c_0>4c$. For $2H<|k|<c_0/\varepsilon$ and throughout the line segment from $0$ to $h$,
\begin{equation}
 |\partial_h^j(I_{n,\varepsilon}-I_{n,0})(k,h)|\le
 C_j\varepsilon |k|^{1-n-j},\qquad 0\le j\le d+2 .
 \label{eq:jetsymbolerror}
\end{equation}
Indeed, set $\rho=\varepsilon|k|$ and rescale the internal momenta by $|k|$. The arguments of all the free propagators then satisfy $1/2\le|u|\le3/2$. There the free dimensionless overlap operator satisfies $D_{\rm dimless}(\rho u)/\rho=i\gamma\cdot u+O_{C^j}(\rho)$, and the same order of error holds for its inverse. The linear gauge vertex differs from its continuum value by $O_{C^j}(\rho)$ in these variables, by the contour and edge integral estimates. In a diagram with $m<n$ propagators the explicit factor $\varepsilon^{n-m}$ from the contact vertices becomes $\rho^{n-m}$ once the common power $|k|^{-n}$ has been extracted. Differentiation in an external momentum gives a further factor $|k|^{-1}$. For the Taylor coefficients at $h=0$ the resulting bounds hold for all $0<|k|<c_0/\varepsilon$.

One extra order of Taylor subtraction removes an apparent ultraviolet logarithm. Let $j_*=5-n=d+1$. Under simultaneous inversion of all four Euclidean coordinates, the two integrands obey
\begin{align}
 I_{n,\varepsilon}(-k,-h)=(-1)^n I_{n,\varepsilon}(k,h),\qquad
 I_{n,0}(-k,-h)=(-1)^n I_{n,0}(k,h).
\end{align}
To verify the lattice transformation, let $\iota x=-x$ and $(\mathcal I\psi)(x)=\gamma_5\psi(-x)$. For an oriented link on the cover,
\begin{align}
 U_a(\iota^*A)_\mu(x)=U_a(A)_\mu(-x-\varepsilon\hat\mu)^{-1}.
\end{align}
Inversion exchanges the forward and backward covariant translations. Its action on spinors commutes with $P_+$ and preserves the spin structure because each boundary-condition sign equals its inverse. The overlap functional calculus then gives $\mathcal I D_a(A)\mathcal I^{-1}=D_a(\iota^*A)$, with the same covariance for each field derivative. Conjugating the trace by $\mathcal I$ preserves the order of its vertices. Each external one-form gives one minus sign. This proves the identities for both terms of \eqref{eq:reallegs} directly on the cover, where the momenta in the Taylor expansion are continuous.

The coefficient of degree $j_*$ has parity $(-1)^{n+j_*}=-1$ in $k$. Its sum vanishes on the shifted grid, and also when restricted to a ball centred at zero. In the lattice sum the partner of $k$ is the representative of $-k$ in the Brillouin zone. Periodicity identifies momenta on opposite zone faces, and an odd coefficient vanishes at a fixed point of this map. In the continuum the coefficient decays as $|k|^{-5}$ and its sum converges absolutely.

We may therefore subtract the Taylor polynomial through degree $j_*$ without changing the polynomial $P_{n,\varepsilon}$. In the intermediate region \eqref{eq:jetsymbolerror} with $d+2$ derivatives bounds the difference of the resulting remainders by $C\varepsilon H^{d+2}\sum_{|k|>2H}|k|^{-5}\le C'\varepsilon H^{d+1}$ after summation. Above $c_0/\varepsilon$ each remainder is bounded by $CH^{d+2}|k|^{-6}$. Their combined contribution is $O(\varepsilon^2H^{d+2})=O(\varepsilon H^{d+1})$. This includes diagrams crossing a zone face because a shift of length at most $H$ changes the distance to the nearest pole by at most $H$.

In the inner ball $|k|\le2H$, evaluate the vertices at the specified external momenta. All internal momenta have modulus at most $3H$ and at least the free gap. The error in a free propagator is $O(\varepsilon)$ and the error in a linear vertex is $O(\varepsilon H)$. Replacing one factor at a time in the product of $n$ propagators gives $C\varepsilon H$ times the product of their scalar bounds. For $m$ propagators with shifts $|p_i|\le H$, H\"older's inequality gives
\begin{align}
 \sum_{|k|\le2H}\prod_{i=1}^m |k+p_i|^{-1}
 \le C\begin{cases} H^{4-m},&m<4,\\ 1+\log H,&m=4.\end{cases}
\end{align}
For $m<n$ there is an extra factor $\varepsilon^{n-m}$ from the contact vertices, so their contribution is $O(\varepsilon H^{d+1})$. For $j\le d$, the difference of the Taylor terms at order $j$ is bounded on summation by $C\varepsilon H^j\sum_{|k|\le2H}|k|^{1-n-j}\le C'\varepsilon H^{d+1}$. The term of order $j_*$ again sums to zero in this ball. A logarithm of $H$ occurs in this estimate only for four linear vertices, as stated in \eqref{eq:jetbound}.
\end{proof}

To obtain the all-momentum bound in \eqref{eq:alljets}, take $2\le n\le4$ and $d=4-n$. The lattice amplitude conserves momentum modulo the reciprocal lattice. For $\varepsilon H<c$ only ordinary conservation, $\sum_i h_i=0$, is possible for a nonzero amplitude. When $\varepsilon H\ge c$ and $\sum_i h_i=0$, at least two legs have momenta comparable to $H$ for large $H$. The lattice vertex is bounded by $C\varepsilon^{-d}\ell_\varepsilon$, while the continuum vertex and counterterm are bounded by $CH^d(1+\log H+\ell_\varepsilon)$. Since $1/\varepsilon\le C H$ and $d\le2$, all are bounded by $C\varepsilon\prod_i(1+|h_i|)^3$. When $\sum_i h_i$ is a nonzero reciprocal-lattice vector, only the lattice vertex contributes. Its bound is $C\varepsilon^{-d}$ for $n=2,3$, by the sums with at most three propagators, and is $C\ell_\varepsilon$ for $n=4$. Each bound is at most $C\varepsilon H^3$, even when a single leg carries the large momentum. This proves the second line of \eqref{eq:alljets} since $H^3\le C\prod_i(1+|h_i|)^3$. Summing the external Fourier amplitudes gives $O(\varepsilon)$ for fields bounded in $\mathcal W_3$. For a band $|h_i|\le M$ with $\varepsilon M$ small, the three- and four-point errors are bounded by $C\varepsilon M^2$ times the unweighted norms. Appendix~\ref{app:vacuum} supplies the two-point bound.

\section{Inversion symmetry and uniform phase matching}
\label{app:odd}

Inversion of all four coordinates removes the logarithmic loss that a factor-by-factor estimate of the five-point function would incur, since estimating the propagator and vertex differences separately leaves a sum of $|k|^{-4}$ over loop momenta. The cancellation takes place in the complex trace, so it controls both the phase and the part of the log modulus left after the four lowest orders have been removed.

\begin{lemma}\label{lem:projectedfive}
Take five plane waves of unit polarization on the torus. Denote their momenta by $h_1,\ldots,h_5$ and take the fifth wave as the variation. Let $\mathcal P_{5,\varepsilon}(h)$ be their multilinear coefficient in the dynamical trace in \eqref{eq:ov} at the vacuum, with all lattice contact vertices included. Let $\mathcal P_{5,0}(h)$ be the corresponding continuum coefficient at the original gauge potential, with the chirality in \eqref{eq:orientation}. Put $H=1+\sum_i|h_i|$. Then
\begin{equation}
 \big|\mathcal P_{5,\varepsilon}(h)-\mathcal P_{5,0}(h)\big|
     \le C\min(1,\varepsilon H).
 \label{eq:fivemodes}
\end{equation}
Both coefficients vanish when all five momenta are zero.
\end{lemma}

\begin{proof}
Use loop momentum $k$ at the output of the varied vertex and strip off the overall momentum conservation factor. Write $J_{5,\varepsilon}(k,h)$ and $J_{5,0}(k,h)$ for the resulting integrands. Each is a finite sum over ordered attachments of the five fields with output projector $P_+$. The continuum weight residue in \eqref{eq:weight} has degree at most four, so only the loop with five linear vertices contributes in the continuum. The lattice trace conserves momentum modulo a reciprocal lattice vector. For $\varepsilon H<c$, with $c>0$ sufficiently small, a nonzero coefficient in either theory requires $\sum_i h_i=0$.

At zero external momentum the spin representative $\gamma_5$ of four-axis inversion commutes with $P_+$. Each constant external one-form, including the variation, changes sign. The covariance of the overlap operator and its field derivatives under this inversion, proved in appendix~\ref{app:localjets}, therefore gives $J_{5,\varepsilon}(-k,0)=-J_{5,\varepsilon}(k,0)$. The continuum integrand has the same parity. The shifted fermion grid is invariant under $k\mapsto-k$, with opposite faces identified on the lattice. Fixed points of the inversion give zero. The continuum sum is absolutely convergent, so the zero-momentum sum vanishes in each theory.

Suppose now that $\varepsilon H<c$ and $\sum_i h_i=0$. Subtract $J_{5,\varepsilon}(k,0)$ and $J_{5,0}(k,0)$ from their respective integrands. To compare the subtracted sums, consider first the annulus $2H<r=|k|<c_0/\varepsilon$, with $4c<c_0$ and $c_0$ small. The routed momenta along the segment from $0$ to $h$ are $k+s p$, $0\le s\le1$, where $p$ is a partial sum of external momenta. Their lengths divided by $r$ lie between $1/2$ and $3/2$. On the cover, the vertices are smooth functions of the independent external momenta with $h_5=-\sum_{j<5}h_j$. The momentum derivative below acts on these functions before the allowed torus momenta are inserted.

Write $\sigma=\varepsilon r$. On the rescaled annulus the free propagators $r S_a$ and $rS$ differ by $O(\sigma)$, also after one derivative in the rescaled momenta. This follows by inverting $D_{\rm dimless}(\sigma u)/\sigma=i\gamma\cdot u+O_{C^1}(\sigma)$ there. For a linear vertex the fixed Wilson contour and the exact edge integral give the same $O_{C^1}(\sigma)$ difference from $\gamma^\mu R_*(A_\mu)$ for a unit external polarization. Hence, after extracting $r^{-5}$, the difference of the five-propagator integrands is $O_{C^1}(\sigma)$. A contact diagram with $m<5$ propagators has the factor $\varepsilon^{5-m}r^{-m}=r^{-5}\sigma^{5-m}$. The remaining dimensionless factors and their first derivatives are bounded by the same contour and edge formulas. Multiplication by $P_+$ preserves these bounds. One derivative with respect to an external momentum component $h_j^\mu$ supplies a factor $r^{-1}$, giving, along the entire segment,
\begin{equation}
 \big|\partial_{h_j^\mu}(J_{5,\varepsilon}-J_{5,0})(k,sh)\big|
       \le C r^{-6}\sigma=C\varepsilon r^{-5},\qquad 1\le j\le4 .
 \label{eq:fiveprojectedderiv}
\end{equation}
The subtracted difference on the annulus thus sums to at most
$C\varepsilon H\sum_{|k|>2H}|k|^{-5}\le C'\varepsilon$.
For $|k|\ge c_0/\varepsilon$, all the routed lattice propagators are $O(\varepsilon)$, since their momenta remain a distance at least $(c_0-c)/\varepsilon$ from the nearest pole. Each lattice diagram is $O(\varepsilon^5)$ on $O(\varepsilon^{-4})$ modes. The continuum tail sums to $O(\varepsilon)$ by its $|k|^{-5}$ decay. These estimates also apply at $h=0$.

On $|k|\le2H$ evaluate the diagrams at the two endpoints separately. All routed momenta there belong to the gapped fermion grid. Replacing a linear vertex costs $O(\varepsilon H)$ times a sum of five shifted propagator bounds, uniformly bounded by H\"older's inequality with exponent five. Replacing a propagator costs $O(\varepsilon)$ times a sum of four such bounds, at most $C(1+\log H)$. This last estimate also covers the four-propagator contact diagrams. A contact diagram with $m<4$ propagators contributes at most $C\varepsilon^{5-m}H^{4-m}$, by summing the $m$-th powers on a ball of radius $3H$. In total the inner region contributes $O(\varepsilon H)$.

Finally, if $\varepsilon H\ge c$, the free $\mathcal S_5$ bound controls the continuum loop and the five-propagator lattice diagrams by a constant. Each lattice contact diagram is bounded by $C\varepsilon^{5-m}\|S_a\|_m^m$, at most $C\varepsilon\ell_\varepsilon$, including when the total momentum is a nonzero reciprocal lattice vector. Since $\min(1,\varepsilon H)\ge\min(1,c)$ in this range, these estimates complete \eqref{eq:fivemodes}.
\end{proof}

The five-point bound can now be summed with the weight $\omega_\varepsilon$ defined in \eqref{eq:fouriermodulus}. Write $f_0=\|f\|_{\mathcal W_0}$. When $A_0\le B$, the sum of the five-point errors is at most $C_B[\omega_\varepsilon(b)+b_0\omega_\varepsilon(A)]$. Indeed, the weight in \eqref{eq:fivemodes} is at most the sum of the five individual weights in \eqref{eq:fouriermodulus}. Only one external field has to carry this weight in each term.

For the remaining background dependence we use the operator bounds \eqref{eq:modop} and the reconstruction bounds \eqref{eq:modrec}. Under the two smallness conditions of theorem~\ref{thm:main} the corresponding inverses are uniformly bounded, as shown in sections~\ref{sec:conv} and~\ref{sec:match}.

Expand the inverse in \eqref{eq:split} by one more term. The trace remainder contains $J$, five factors of $K$, and a bounded inverse. In replacing one of the six propagator factors, take its difference in operator norm and the other five in $\mathcal S_5$. If the inverse is replaced, its difference is $O_{B,G}(\omega_\varepsilon(A))$ by the resolvent identity, and five of the remaining six factors suffice for the trace. These replacements cost at most $C_{B,G}[\omega_\varepsilon(b)+b_0\omega_\varepsilon(A)]$. The same argument with $V(A^r-A)S$ and $V(b^r-b)S$ applies to the change of the continuum remainder under reconstruction. The continuum five-point coefficient changes by at most the same amount since its five propagators already give a trace-class product.

The finite polynomial from the inverse expansion also contains lattice contact terms of degree $n\ge6$ with $m\le5$ propagators. At fixed $m$ their absolute trace bound contains $\varepsilon^{n-m}\|S_a\|_m^m$ times the vertex bounds after their powers of $\varepsilon$ have been extracted. For $\varepsilon$ sufficiently small at fixed $B$, the geometric series from \eqref{eq:vertexbd} absorbs the polynomial number of ordered attachments at each $n$. The sum for $m=5$ is $O_B(\varepsilon b_0)$. For $m=4$ it is $O_B(\varepsilon^2\ell_\varepsilon b_0)$, and for $m<4$ it is $O_B(\varepsilon^2 b_0)$. The exact measure term is $O_B(\varepsilon b_0)$ by \eqref{eq:measure}.

Combining these bounds with lemma~\ref{lem:projectedfive} bounds the full complex variation after removal of its low-order terms. Explicitly,
\begin{equation}
 \left|\delta_b\log\frac{W_{\mathrm{ov}}^{(a)}}{W_{\mathrm{CW}}^{(a)}}-L_a(A;b)+L_c(A^r;b^r)\right|
 \le C_{B,G}\big[\omega_\varepsilon(b)+b_0\omega_\varepsilon(A)\big].
 \label{eq:phasetailmodulus}
\end{equation}
The two polynomials are real, so taking the imaginary part proves \eqref{eq:phasefine}. The bound $\omega_\varepsilon(f)\le\varepsilon^\alpha\|f\|_{\mathcal W_\alpha}$ for $0<\alpha\le1$ then gives \eqref{eq:main}, including its endpoint. For fields with Fourier support $|h|\le M$, one has $\omega_\varepsilon(f)\le\varepsilon(1+M) f_0$, which gives the high-order estimate used in corollary~\ref{cor:band}.

\section{Fixed-order convergence at fixed external momenta}
\label{app:fixedfive}

At fixed external momenta, step reversal improves the convergence of each Taylor coefficient. We first establish the five-point estimate, including the measure and reconstruction contributions to the phase, and then extend the argument to every fixed order. These estimates do not control the sum over orders at a background of fixed amplitude.

\begin{lemma}\label{lem:fixedfive}
Fix five torus momenta $h=(h_1,\ldots,h_5)$. With the coefficients and chirality of lemma~\ref{lem:projectedfive}, there are $C_h,\varepsilon_h>0$ such that, for unit external polarizations and $0<\varepsilon<\varepsilon_h$,
\begin{equation}
 \big|\mathcal P_{5,\varepsilon}(h)-\mathcal P_{5,0}(h)\big|
       \le C_h\varepsilon^2\ell_\varepsilon .
 \label{eq:fixedfive}
\end{equation}
The comparison is at the original gauge potential and includes every dynamical contact vertex at the vacuum.
\end{lemma}

\begin{proof}
For the signed displacement $t$ of appendix~\ref{app:vacuum}, differentiating \eqref{eq:signedstep} in real sources and then extending multilinearly gives
\begin{align}
 D_{-t}=-\gamma_5D_t\gamma_5,\qquad
 S_{-t}=-\gamma_5S_t\gamma_5,\qquad
 V_{q,-t}=-\gamma_5V_{q,t}\gamma_5 .
\end{align}
Here $V_{q,t}$ is a complete $q$-field vertex. An ordered graph is a trace of factors $V_{q,t}S_t$ with $P_+$. Step reversal conjugates the product by $\gamma_5$, which commutes with $P_+$; the graph is therefore even in $t$. This statement concerns complete contact vertices, not their separate algebraic terms.

Put $H=1+\sum_i|h_i|$. A fixed nonzero $\sum_i h_i$ cannot be a reciprocal-lattice vector for sufficiently small $\varepsilon$, so both coefficients vanish unless $\sum_i h_i=0$. In that case use the kernels $J_{5,t}(k,h)$ of lemma~\ref{lem:projectedfive}, with $h_5=-\sum_{j<5}h_j$. Four-axis inversion makes their zero-external-momentum sums vanish separately. Subtract $J_{5,t}(k,0)$ in each entire sum before splitting the loop momenta into regions.

On $2H<r=|k|<c_0/\varepsilon$, all routed momenta along the segment from $0$ to $h$, divided by $r$, have lengths between $1/2$ and $3/2$. Set $\sigma=tr$. A graph with $m$ propagators has the factor
\begin{align}
 t^{5-m}r^{-m}=r^{-5}\sigma^{5-m}.
\end{align}
The fixed Wilson contour and exact edge integrals give smooth functions of $\sigma$ and the rescaled momenta on this compact annulus. The free inverse is regular there, and the apparent singularity at $\sigma=0$ is removable. The entire rescaled graph, including $\sigma^{5-m}$, has uniformly bounded mixed derivatives up to two in $\sigma$ and one in the external momenta divided by $r$. It is even in $\sigma$, with its value at zero equal to the continuum graph. Thus
\begin{equation}
 \big|\partial_{h_j^\mu}(J_{5,\varepsilon}-J_{5,0})(k,sh)\big|   \le C r^{-6}(\varepsilon r)^2=C\varepsilon^2r^{-4},
 \qquad 0\le s\le1,\quad j<5 .
 \label{eq:fixedfiveannulus}
\end{equation}
Integration along the segment and summation over four-dimensional shells give
\begin{align}
 C\varepsilon^2H\sum_{2H<|k|<c_0/\varepsilon}|k|^{-4}   \le C_h\varepsilon^2\ell_\varepsilon .
\end{align}

The inner region $|k|\le2H$ has finitely many modes. Evaluate the kernels only at $h$ and $0$; their routed momenta lie in the fixed gapped fermion grid. Smoothness and evenness in $t$ give $O_h(\varepsilon^2)$ after summation, without interpolating momenta through this region.

Finally, choose $\varepsilon H<c_0/4$. Above $c_0/\varepsilon$, routed lattice momenta stay a distance of order $1/\varepsilon$ from the nearest pole along the whole segment. Each graph is $\varepsilon^5$ times a smooth function of $\varepsilon k,\varepsilon h$; an external derivative supplies another $\varepsilon$. The subtracted graph is therefore $O(H\varepsilon^6)$, whose sum over $O(\varepsilon^{-4})$ modes is $O_h(\varepsilon^2)$. Periodicity gives the same bound across Brillouin-zone faces. The corresponding continuum derivative is $O(r^{-6})$, so its tail has the same order. This proves \eqref{eq:fixedfive}.
\end{proof}

Both symmetries are needed: step reversal alone gives a middle-region difference $O(\varepsilon^2r^{-3})$, whose absolute sum is only $O(\varepsilon)$. The zero-momentum subtraction supplies the extra external derivative in \eqref{eq:fixedfiveannulus}.

\begin{corollary}\label{cor:fixedfivephase}
Fix five plane waves $f_i$ with momenta $h_i$ and unit polarizations. Put $A(s)=\sum_{i=1}^4s_i f_i$ and $b=f_5$. Then
\begin{equation}
 \left|\left.\partial_{s_1}\cdots\partial_{s_4}
 \left(\operatorname{Im}\,\delta_b\log
 \frac{W_{\mathrm{ov}}^{(a)}[U_a(A(s))]}{W_{\mathrm{CW}}^{(a)}[U_a(A(s))]}\right)
 \right|_{s=0}\right|
 \le C_h\varepsilon^2\ell_\varepsilon .
 \label{eq:fixedfivephase}
\end{equation}
The variation supplies the fifth field. The derivatives are defined first for real fields and then extended multilinearly to Fourier modes.
\end{corollary}
\begin{proof}
Choose a fixed small ball of real source amplitudes spanning the real and imaginary parts of the five waves. On this smooth family, the estimate in app.\,F.3 of \cite{Craig:2026soq} bounds the exact dimensionless current and each fixed source derivative by $O_h(\varepsilon^9)$. Returning from its midpoint frame uses short parallel transports with bounded source derivatives and preserves the estimate. The link variation and its source derivatives are $O_h(\varepsilon)$. Taking four source derivatives of their contraction and summing over $O(\varepsilon^{-4})$ links gives an $O_h(\varepsilon^6)$ measure coefficient. The parameter derivatives in the cited estimate are essential; a pointwise asymptotic bound would not suffice.

For reconstruction, let $\Phi=\arg W_{\rm cont}$ near the vacuum and $R_{1,\varepsilon}=dA^r(0)$. Section~\ref{sec:response} gives $d^j\Phi(0)=0$ for $1\le j\le4$, while $A^r(0)=0$. The chain rule therefore gives exactly
\begin{align}
 d^5(\Phi\circ A^r)(0)[f_1,\ldots,f_5] =d^5\Phi(0)[R_{1,\varepsilon}f_1,\ldots,R_{1,\varepsilon}f_5].
\end{align}
This includes the varied fifth leg. Lemma~\ref{lem:affine} bounds $\|(R_{1,\varepsilon}-1)f_i\|_\infty$ by $C_h\varepsilon^2$. The fifth continuum coefficient is a sum of traces with five factors $V(f)S$, each bounded in $\mathcal S_5$ by $C\|f\|_\infty$. Replacing the five legs in turn thus costs $O_h(\varepsilon^2)$, including every Fourier component generated by reconstruction. Combining this with the measure estimate and lemma~\ref{lem:fixedfive} proves the result.
\end{proof}

The argument extends to every fixed order of the complete complex response.

\begin{proposition}\label{prop:allorders}
Fix $n\ge2$ and torus momenta $h_1,\ldots,h_n$ with $\sum_i h_i=0$. Let $\mathcal D_{n,a}(h)$ be the $n$-linear Taylor coefficient at the vacuum, on unit polarizations, of
\begin{align}
 \delta_b\log\frac{W_{\mathrm{ov}}^{(a)}[U_a(A)]}{W_{\mathrm{CW}}^{(a)}[U_a(A)]}   -c_R(a,\mu)\,\delta_bI(A),
\end{align}
counting the variation as one field. There are $C_{n,h},\varepsilon_{n,h}>0$ such that
$ |\mathcal D_{n,a}(h)|\le C_{n,h}\varepsilon^2\ell_\varepsilon$ for $0<\varepsilon<\varepsilon_{n,h}$.
\end{proposition}
\begin{proof}
First compare the dynamical coefficients at the original field. The step-reversal argument above holds for any fixed number of fields; the contour formulas provide any fixed number of external-momentum derivatives on the rescaled annulus. For $n\le5$, subtract the external-momentum Taylor polynomial through degree $5-n$ in each regulator. By the inversion identity of appendix~\ref{app:localjets}, the term of degree $5-n$ is odd in the loop momentum and sums to zero. This cancellation is made in each complete symmetric momentum sum before the split into regions; the continuum odd term decays as $|k|^{-5}$ and is absolutely summable. Terms of degree at most $4-n$ form the local polynomial, with the spectral prescription of appendix~\ref{app:vacuum}. Taylor's formula and step evenness give, for the subtracted kernels on $2H<r<c_0/\varepsilon$,
\begin{align}
 |J_{n,\varepsilon}^{\rm sub}-J_{n,0}^{\rm sub}|  \le C_n\varepsilon^2 H^{6-n}r^{-4}.
\end{align}
For $n\ge6$ no subtraction is needed: a graph with $m$ propagators is $r^{-n}$ times a smooth function containing $(\varepsilon r)^{n-m}$. Its difference from the continuum graph is $O_n(\varepsilon^2r^{2-n})$. Four-dimensional shell sums give $O_{n,h}(\varepsilon^2\ell_\varepsilon)$, with a logarithm only at $n=6$ in the latter case. The finite inner region is treated at the endpoints as above. In the outer region, $n\le5$ requires $6-n$ external derivatives and hence gives $\varepsilon^6$ per mode; for $n\ge6$ the bound is $\varepsilon^n$. Both lattice sums and continuum tails are $O_{n,h}(\varepsilon^2)$.

At $n=2$, lemma~\ref{lem:evenbubble} already fixes the local polynomial to this accuracy after subtraction of $c_RI$. For $n=3,4$, apply the fixed finite-dimensional left inverse in the proof of theorem~\ref{thm:ward}, now using the improved remainders just obtained. The Ward identities successively bound their residual local coefficients by $O(\varepsilon^2\ell_\varepsilon)$. The measure contributes $O_{n,h}(\varepsilon^6)$: use the same smooth source ball as in corollary~\ref{cor:fixedfivephase}, with $n-1$ parameter derivatives.

It remains to control reconstruction at order $n$. For a block $B$ of the prescribed input fields, put $h_B=\sum_{i\in B}h_i$. The Taylor coefficient of $E(A)=A^r(A)-A$ on that block has the form
\begin{equation}
 E_B(x)=\sum_{m\in\mathbb Z^4}e_B(m)\, e^{i(h_B+2\pi m/\varepsilon)\cdot x},\qquad
 |e_B(m)|\le C_{n,h,J}\varepsilon^2(1+|m|)^{-J}
 \label{eq:fixedalias}
\end{equation}
for every fixed $J$. This is the reference-cell estimate in the proof of lemma~\ref{lem:foldedrecon}: affine reproduction handles a one-field block, the constant-field cancellation handles two fields, and \eqref{eq:cellcoefficient} gives $\varepsilon^{|B|-1}\le\varepsilon^2$ for larger blocks. Smooth matching across cell faces permits integration by parts in the periodic cell variable to obtain the displayed decay.

The chain rule gives finitely many partitions of the $n$ inputs among continuum vertices. At vertices of degree at least five, continuity on $\mathcal W_0$ bounds every term containing an error by $O_{n,h}(\varepsilon^2)$. For vertices of degree two through four, the kernel is bounded by $CH^2(1+\log H)$. If precisely one argument is an error coefficient, momentum conservation forces its alias index to be zero, since $\sum_i h_i=0$ and all other arguments are original waves. The kernel then has fixed momenta and the term is $O_{n,h}(\varepsilon^2)$. With $r\ge2$ error arguments, \eqref{eq:fixedalias} bounds the alias sum by a constant times
\begin{align}
 \varepsilon^{2r-2}\ell_\varepsilon
 \sum_{m_1,\ldots,m_r}
 \frac{(1+\sum_i|m_i|)^2[1+\log(2+\sum_i|m_i|)]}  {\prod_i(1+|m_i|)^J} =O_{n,h}(\varepsilon^2\ell_\varepsilon),\qquad J>7 .
\end{align}
We have dropped the momentum-conservation restriction to bound the sum absolutely. This accounts also for nonlinear reconstruction terms entering low-point vertices. The one-point vertex vanishes. Adding the dynamical, measure and reconstruction bounds proves the proposition.
\end{proof}

The constants may grow with both the order and the external momenta. Thus the proposition proves a fixed-order bound, with a logarithm for both real and imaginary parts; it neither sums the background expansion nor establishes a log-free phase estimate. The uniform bounds of theorems~\ref{thm:main} and~\ref{thm:finite}, and the bandwidth window in corollary~\ref{cor:band}, are unchanged.

\section{Estimates for the collar reconstruction}
\label{app:collar}

The phase comparison uses the reconstructed potential. For nonzero Fourier modes, the quadratic modulus can be expressed through its linearized curvature; the zero-mode holonomy response is treated separately. We first bound the potential and curvature, then separate the universal short-distance part of the continuum two-point function from its finite-volume remainder.

The linearization \eqref{eq:linearrec} uses only the partition of unity and $X_{ji}=-X_{ij}$, and therefore remains valid:
\begin{equation}
 R^{\rm c}u=\sum_{ij}\rho_i\,d\rho_j\,X_{ij}(u),\qquad \rho_i=\alpha_i^2.
 \label{eq:collarlin}
\end{equation}
For $u=d\lambda$, it equals $d(\sum_j\rho_j\lambda(x_j))$. In particular, a constant potential $c$ is mapped to $\sum_\mu c_\mu p(x_\mu/\varepsilon)dx^\mu$.

\begin{lemma}\label{lem:collarpot}
For $\|A\|_{\mathcal W_1}\le B$, the collar potential and its variation, in the periodic frame constructed as in lemma~\ref{lem:reconstruction} with the collar weights, satisfy
\begin{align}
 \|A^r_\mu-p(x_\mu/\varepsilon)A_\mu\|_\infty\le C_B\varepsilon,
 \qquad
 \|\delta_b A^r_\mu-p(x_\mu/\varepsilon)b_\mu\|_\infty
 \le C_B\varepsilon\|b\|_{\mathcal W_1}.
\end{align}
For a fixed wave $u(x)=v e^{ihx}$, let $f_{\mu\nu}=i(h_\mu v_\nu-h_\nu v_\mu)$. Then
\begin{equation}
 [d(R^{\rm c}u)]_{\mu\nu}
 =e^{ihx}\{p(y_\mu)p(y_\nu)f_{\mu\nu}+e_{\mu\nu,\varepsilon}(y;h,v)\},
 \quad y=x/\varepsilon,
 \label{eq:collarcell}
\end{equation}
where $e_\varepsilon$ is periodic in $y$ and, for every fixed $J$,
\begin{equation}
 \|e_\varepsilon\|_{C_y^J}\le C_{J,h}\varepsilon|v|,
 \qquad
 \sum_{n\in\mathbb Z^4}[1+\log(2+|n|)]|\widehat e_\varepsilon(n)|^2
 \le C_h\varepsilon^2|v|^2.
 \label{eq:collaralias}
\end{equation}
The constants may depend on the fixed profile.
\end{lemma}
\begin{proof}
The smooth cell map and its field derivative have the bounds used in \eqref{eq:cellcoefficient}. Its nonlinear terms contribute $O_B(\varepsilon)$ to the potential, including one field derivative. In the linear term, replacing the field on the contributing edges by its value at the cell centre changes \eqref{eq:collarlin} by $O(\varepsilon\|\nabla u\|_\infty)$; the constant part gives $p(y_\mu)u_\mu$. This proves the potential bounds.

To compute the linear curvature, take $u=f x^\mu dx^\nu$, $\mu\ne\nu$, in a cell based at zero. Averaging the two orders of the $\mu$- and $\nu$-segments assigns half the flux to either diagonal path. Substitution in \eqref{eq:collarlin} gives
\begin{align}
 (R^{\rm c}u)_\nu
 =f\varepsilon\sin^2\vartheta(y_\mu)p(y_\nu),\qquad
 (R^{\rm c}u)_\mu=0.
\end{align}
The other weights sum to one. The resulting curvature is $f p(y_\mu)p(y_\nu)$. Constant and exact affine one-forms have zero curvature, so this computation determines the result for every affine input.

Translation covariance makes $e^{-ihx}d(R^{\rm c}u)$ periodic with period $\varepsilon$. It is a smooth cell function, analytic in $q=\varepsilon h$, obtained from finitely many edge integrals of the wave. Before the factor $\varepsilon^{-1}$ associated with its curvature is restored, its constant term in $q$ vanishes and its linear term is given by the affine calculation above. Taylor's formula leaves $O_J(|q|^2)$ in every fixed cell norm $C_y^J$. Thus the curvature remainder is $O_{J,h}(\varepsilon)|v|$, including after removal of the Bloch factor. Smoothness across cell boundaries follows from that of the weights. Integration by parts on the reference torus gives
$ |\widehat e_\varepsilon(n)|\le C_{J,h}\varepsilon|v|(1+|n|)^{-J}$;
choosing $J>2$ proves the weighted sum in \eqref{eq:collaralias}. This is an additive estimate and requires no division by a profile Fourier coefficient.
\end{proof}

\begin{lemma}\label{lem:toruspolarization}
Let $\mathcal P_T(k)$ and $\mathcal P_\infty(k)$ be the vacuum Hessian tensors of the continuum log modulus on the unit torus and on $\mathbb R^4$, in the same spectral prescription and at the same scale. At torus momenta,
\begin{equation}
 \|\mathcal P_T(k)-\mathcal P_\infty(k)\|\le C,
 \label{eq:toruspolarization}
\end{equation}
with $\mathcal P_\infty(0)=0$. Both tensors are transverse for $k\ne0$; no rotational invariance of $\mathcal P_T$ is asserted.
\end{lemma}
\begin{proof}
Let $S_T(x)$ be the free torus propagator with the prescribed spin signs. Near the diagonal,
$S_T(x)=S_\infty(x)+H(x)$, where $H$ is smooth: both kernels solve the same local Dirac equation with the same delta source, and their difference has no singularity by elliptic regularity. Here $S_\infty(x)=O(|x|^{-3})$.

The finite contact terms are fixed by the same spectral prescription: with $Q_0=-D_0^2$, differentiating $\mathcal M=-\zeta'_{Q/\bar\mu^2}(0)/4$ twice gives
\begin{align}
 \delta_v\delta_u\mathcal M(0)
 =-\tfrac12\FP_{z=0}\Tr\!\left[V(u)S V(v)S
                    (Q_0/\bar\mu^2)^{-z}\right]
   +\mathcal C_{\bar\mu}(u,v),
\end{align}
where $\mathcal C_{\bar\mu}$ consists of the local residues from differentiating the spectral weight and cyclically arranging the factors, as in \eqref{eq:weight}. These residues agree on the two spaces because their complete local Dirac symbols and spectral scale coincide. The weighted propagator likewise has a common singular part $S_{\infty,z}(x)=O(|x|^{-3+2\operatorname{Re} z})$ and a smooth remainder, so the difference of the weighted bubbles is locally integrable and holomorphic for $|\operatorname{Re} z|<1/4$; its finite part is its ordinary value at $z=0$, with no residual delta distribution or derivative thereof.

Periodize the infinite-volume Hessian distribution. Away from zero its kernel and its derivatives decay as $|x|^{-6}$ and higher powers, so the nonzero translates converge smoothly near the diagonal. The difference between the torus kernel and this periodized distribution is consequently an $L^1$ matrix kernel on the torus. Its Fourier coefficients are uniformly bounded, proving \eqref{eq:toruspolarization}. Gauge invariance gives the stated transversality. The zero Fourier coefficient of the remainder may be nonzero; it includes the holonomy response discussed in appendix~\ref{app:vacuum}.
\end{proof}

\begin{proof}[Proof of theorem~\ref{thm:collar}]
(a) Let $P$ be the zero-mean periodic primitive of $p-1$, and set
\begin{align}
 g=\exp\!\left[-\varepsilon\sum_\mu P(x_\mu/\varepsilon)A_\mu(x)\right].
\end{align}
Lemma~\ref{lem:collarpot} gives
$g^{-1}A^rg+g^{-1}dg=A+O_B(\varepsilon)$ in maximum norm, also after one field derivative, with error $C_B\varepsilon\|b\|_{\mathcal W_1}$. Gauge invariance, \eqref{eq:pullback} and the comparison at the original field in theorem~\ref{thm:main} prove \eqref{eq:collarphase}. The same small perturbation preserves the Dirac gap.

(b) Put $\mathcal M_2(u)=[s^2]\mathcal M(su)$. The one-point function vanishes at the vacuum, so only $R^{\rm c}$ enters the reconstructed quadratic coefficient. For one Fourier pair $\pm h$, with coefficient $v$ at $h$, write the output momenta as $k_n=h+2\pi n/\varepsilon$. At $n=0$, the potential coefficient is $v+O_h(\varepsilon)|v|$, since $\int p=1$; its contribution to the difference of quadratic forms is $O_h(\varepsilon)|v|^2$. For $h=0$, it agrees exactly: $R^{\rm c}c-c=d[\varepsilon\sum_\mu c_\mu P(x_\mu/\varepsilon)]$, and the vacuum Ward identity removes this periodic gradient. Thus no assumption that the torus holonomy response vanishes is needed.

For $n\ne0$, only the transverse potential contributes, and its norm is bounded by the linear curvature divided by $|k_n|$. Lemma~\ref{lem:toruspolarization} and \eqref{eq:collaralias} therefore bound the sum of finite-volume corrections by $O_h(\varepsilon^2)|v|^2$. It remains to use the infinite-volume scalar coefficient \eqref{eq:flatquillen}. The leading curvature coefficients in the $\mu\nu$ plane are
\begin{align}
 \hat p_{n_\mu}\hat p_{n_\nu}
 \prod_{\lambda\ne\mu,\nu}\mathbf1_{n_\lambda=0}\,f_{\mu\nu}.
\end{align}
Parseval gives a total nonzero-mode energy of $(\kappa-1)I_2(u)$, and its sum weighted by $\log(4\pi^2|n|^2)$ is $\Xi I_2(u)$. On these modes
\begin{align}
 \log\frac{|k_n|^2}{\bar\mu^2}
 =\log(4\pi^2|n|^2)-\log(\varepsilon^2\bar\mu^2)
   +O_h\!\left(\frac\varepsilon{|n|}\right).
\end{align}
The additive remainder in \eqref{eq:collaralias}, paired with the rapidly decreasing leading coefficients by Cauchy-Schwarz, changes the weighted energy by $O_h(\varepsilon\ell_\varepsilon)|v|^2$. The remainder squared is smaller. Distinct momenta in the fixed input support cannot differ by a nonzero reciprocal lattice vector for sufficiently small $\varepsilon$, so summing the finitely many pairs yields
\begin{align}
 \mathcal M_2(R^{\rm c}u)-\mathcal M_2(u)
 =\frac{T(R)}{24\pi^2}
 \left[\Xi-(\kappa-1)\left(\log(\varepsilon^2\bar\mu^2)+\frac53\right)\right]I_2(u)
 +O(\varepsilon\ell_\varepsilon).
\end{align}
Subtracting this expression from theorem~\ref{thm:two}, and using $\varepsilon\bar\mu=a\mu$, gives \eqref{eq:collarcR}.
\end{proof}

\providecommand{\href}[2]{#2}\begingroup\raggedright\endgroup

\end{document}